\documentclass[journal, letterpaper]{IEEEtran}
\IEEEoverridecommandlockouts
\usepackage{ifpdf}
\normalsize
\usepackage{cite}
\usepackage[pdftex]{graphicx}
\DeclareGraphicsExtensions{{.pdf}}
\usepackage{units, nicefrac}
\usepackage[printonlyused]{acronym}
\usepackage[cmex10]{amsmath}
\usepackage{bigints}
\usepackage{mathtools,amssymb}
\usepackage{dsfont}
\usepackage{amsthm}
\usepackage{mdwmath}
\usepackage{mdwtab}
\usepackage[tight,hang,footnotesize]{subfigure}
\usepackage{stfloats}
\usepackage{color}
\usepackage{multirow}
\usepackage{comment}
\usepackage{longtable}
\fnbelowfloat

\DeclareFontFamily{U}  {MnSymbolB}{}
\DeclareFontShape{U}{MnSymbolB}{m}{n}{
    <-6>  MnSymbolB5
   <6-7>  MnSymbolB6
   <7-8>  MnSymbolB7
   <8-9>  MnSymbolB8
   <9-10> MnSymbolB9
  <10-12> MnSymbolB10
  <12->   MnSymbolB12}{}

\DeclareFontShape{U}{MnSymbolB}{b}{n}{
    <-6>  MnSymbolB-Bold5
   <6-7>  MnSymbolB-Bold6
   <7-8>  MnSymbolB-Bold7
   <8-9>  MnSymbolB-Bold8
   <9-10> MnSymbolB-Bold9
  <10-12> MnSymbolB-Bold10
  <12->   MnSymbolB-Bold12}{}

\DeclareSymbolFont{MnSyB}         {U}  {MnSymbolB}{m}{n}
\SetSymbolFont{MnSyB}       {bold}{U}  {MnSymbolB}{b}{n}
\DeclareMathSymbol{\nuparrow}{\mathrel}{MnSyB}{1}

\newcommand{\Prob}[0]{\mathbb{P}}
\newcommand{\Exp}[0]{\mathbb{E}}

\acrodef{CPRI}[CPRI]{Common Public Radio Interface}
\acrodef{DRAN}[DRAN]{distributed RAN}
\acrodef{FLOP}[FLOP]{floating point operation}
\acrodef{HARQ}[HARQ]{Hybrid Repeat Request}
\acrodef{MCS}[MCS]{Modulation and Coding Scheme}
\acrodef{RAN}[RAN]{Radio Access Network}
\acrodef{SNR}[SNR]{Signal-to-Noise Ratio}

\begin{document}
%

\title{Are Heterogeneous Cloud-Based Radio Access Networks Cost Effective?}
%

\author{Vinay~Suryaprakash, ~\IEEEmembership{Member, ~IEEE,}
        Peter Rost, ~\IEEEmembership{Senior Member, ~IEEE,}
        and~Gerhard~Fettweis, ~\IEEEmembership{Fellow, ~IEEE}
\thanks{V. Suryaprakash and G. Fettweis are with the Vodafone Chair Mobile Communications Systems, Technische Universit\"{a}t Dresden, 01062, Dresden. e-mail: (vinay.suryaprakash, gerhard.fettweis)@ifn.et.tu-dresden.de.}
\thanks{Peter Rost is with NEC Laboratories Europe, Kurf\"{u}rsten-Anlage 36 D-69115 Heidelberg. e-mail: (Peter.Rost)@neclab.eu.}
\thanks{The research leading to these results has received funding from the European Union Seventh Framework Programme (FP7/2007-2013) under grant agreement no. 317941 -- project iJOIN.}.
}
\vspace{-1cm}
\maketitle

\begin{abstract}
Mobile networks of the future are predicted to be much denser than today's networks in order to cater to increasing user demands. In this context, cloud based radio access networks have garnered significant interest as a cost effective solution to the problem of coping with denser networks and providing higher data rates. However, to the best knowledge of the authors, a quantitative analysis of the cost of such networks is yet to be undertaken. This paper develops a theoretic framework that enables computation of the deployment cost of a network (modeled using various spatial point processes) to answer the question posed by the paper's title. Then, the framework obtained is used along with a complexity model, which enables computing the information processing costs of a network, to compare the deployment cost of a cloud based network against that of a traditional LTE network, and to analyze why they are more economical. Using this framework and an exemplary budget, this paper shows that cloud-based radio access networks require approximately 10 to 15\% less capital expenditure per square kilometer than traditional LTE networks. It also demonstrates that the cost savings depend largely on the costs of base stations and the mix of backhaul technologies used to connect base stations with data centers.
\end{abstract}

\IEEEpeerreviewmaketitle

\section{Introduction}
Future mobile network deployments are expected to be much denser than networks of today in order to provide significantly higher data rates to a larger number of users. This densification of networks also necessitates novel technologies which are able to cope with more complex deployment and interference scenarios, and are able to improve the utilization of the network, while providing the flexibility required to adapt to a scenario at hand. In this context, centralization of the \ac{RAN} functionality plays an essential role. In a centralized \ac{RAN}, functionality of the protocol stack is executed at a central data center or cloud platform, which is why we refer to this type of deployment as Cloud-\ac{RAN} in this work. Cloud-\ac{RAN} requires additional infrastructure for data centers while the communication infrastructure is less complex and may be conceivably cheaper, as indicated by \cite{ChinaMobile.CRAN.WhitePaper.2011} and \cite{Guan.Kolding.Merz.CRAN.2010}. Hence, there is an inherent trade-off between improved system performance and the costs incurred in a centralized \ac{RAN}.

\subsection{Cost efficiency in Cloud-RAN}
So far, quantitative studies on Cloud-\ac{RAN} have focused on improvements in terms of throughput and energy efficiency. The constraints under which Cloud-RAN is sustainable from a cost perspective are, therefore, unclear. These constraints, however, will play an important role in the decision to deploy Cloud-RAN. In a Cloud-RAN system, many different \textit{interdependent} factors determine the cost of deployment and operation, such as device intensities, equipment cost, capacity cost, infrastructure cost, and data processing cost. This paper is, to the best of our knowledge, the first complete analysis which takes both communication and data processing capabilities as well as their relationship to deployment costs into account. Furthermore, it also considers how the costs (listed above) are related and how different system parameters impact the overall cost. The framework derived in this paper allows identifying operating regimes in which Cloud-\ac{RAN} proves more cost effective. Most importantly, the framework derived is parametrized in a way that permits the evaluation of various deployments and utilization scenarios.

\subsection{Related Work}
Though knowledge of the total cost of networks and methods of modeling them have always been important, there are not many non-proprietary works which show how the cost (irrespective of whether it is capital expenditure or operational expenditure) of an entire network can be computed. The first paper which addressed this topic was \cite{FB'97} and it demonstrated how the deployment costs of a fixed line telecommunication network could be calculated based on a model using homogeneous Poisson processes. Inspired by which, our paper \cite{ICC'14} used a similar model (using homogeneous Poisson processes) to compute the deployment costs of a homogeneous mobile network including the entire backbone infrastructure. This work was extended upon in \cite{WiOpt'14}, which modeled heterogeneous networks (along with their backhaul infrastructure) using various stationary point processes. The above mentioned works, however, take neither the Cloud-\ac{RAN} concept nor the additional information processing costs into account. Now, as far as Cloud-\ac{RAN} is concerned, since its introduction in \cite{Guan.Kolding.Merz.CRAN.2010}, the concept (as a whole) has drawn significant attention. Not long ago, the Next Generation Mobile Networks Alliance published a technical report (see \cite{NGMN.CRAN.TechReport.2013}) which 
states that, besides performance improvements through multi-cell processing, improvements in cost- and energy-efficiency are also expected. However, the report does not provide a quantitative analysis of the characteristic benefits of a Cloud-RAN system. In comparison, the white paper \cite{ChinaMobile.CRAN.WhitePaper.2011} states that Cloud-\ac{RAN} reduces capital expenditure by $15$\% and operational expenditure by $50$\% when compared to a (traditionally deployed) distributed network. The report, however, does not detail how these numbers were obtained.

Most of the work on Cloud-\ac{RAN} focuses on fully centralized networks, i.e., all \ac{RAN} functionality is executed at the data center. In contrast, \cite{iJOINref'13} proposed a flexible centralization of \ac{RAN} depending on the actual backhaul network characteristics. This flexibility allows exploiting a part of the centralization gain despite non-ideal connections between small-cells and a data center. However, as detailed in \cite{Wuebben.etal.SPM.2014}, this is accompanied by challenges in the operation of Cloud-\ac{RAN} and the signal processing performed. In addition to which, none of the literature available considers the relationship between the data processing resources required and the mobile communication traffic offered. 
In \cite{Bhaumik.Chandrabose.Jataprolu.Kumar.Muralidhar.Polakos.Srinivasan.Woo.MobiCom.2012}, Bhaumik et al. provided a comprehensive quantitative assessment of the computational resources required for a specific configuration of a 3GPP LTE mobile network and showed that the turbo-decoding process requires a majority of the processing resources. However, \cite{Bhaumik.Chandrabose.Jataprolu.Kumar.Muralidhar.Polakos.Srinivasan.Woo.MobiCom.2012} does not provide a model which allows extrapolating these results to different system configurations and nor does it enable quantifying centralization gain in a Cloud-\ac{RAN} system.

The first comprehensive analytical model to assess decoder complexity was by Grover et al. in \cite{Grover.Woyach.Sahai.JSAC.2011}. 
In \cite{Rost.Talarico.Valenti.TWC.2014}, this model was extended to include the entire mobile network and to allow quantifying the data processing requirements. This framework allows to dimension the data processing resources of a Cloud-\ac{RAN} system such that a given quality criterion is met; e.g., the probability that the system has insufficient data processing resources to process all incoming transmissions, which is referred to as computational outage \cite{Valenti.Talarico.Rost.GC.2014}.

\subsection{Contributions and organization of the paper}
This paper introduces a model using various spatial point processes for analyzing the deployment costs of a Cloud-RAN system which takes users, base stations (both macro and micro), backhaul (both microwave and optic fiber), and data centers into account. An expression for the average cost of deploying a data center is derived from which the total deployment cost of the network is found. Then, we utilize a data processing model (which helps dimension the data center based on the traffic demands as well as decoder quality) to provide values which can be used in the expression for deployment cost. Using these values, we examine whether Cloud-RAN based networks are more economical than traditional LTE networks and the reasons behind it. Numerical evaluations reveal that Cloud-RAN based networks are indeed more cost effective with respect to deployment costs because they are better at adapting to network load (i.e., the number of active users) and can exploit the fact that the number of processors required does not increase linearly with the load. However, these evaluations also highlight the fact that deployment cost of Cloud-RAN technologies increases when user and data center intensities increase and when the deployment favors a particular backhaul technology.

The paper is structured as follows. Section \ref{Problem_Framework} introduces our spatial point process system model and the underlying data processing model. Section \ref{Cost_Model} introduces the various costs involved and derives the framework used to analyze the deployment cost of Cloud-\ac{RAN} systems. Section \ref{Numerical_Evaluation} evaluates the findings of Section \ref{Cost_Model} numerically and discusses the quantitative results. The paper is concluded in Section \ref{End}.

\section{Problem Framework}\label{Problem_Framework}
The problem framework consists of two subparts. The first, Section \ref{System_Model}, describes the model which is used to obtain the framework to calculate the deployment cost. Section~\ref{Complexity_Model} then describes the dependence of deployment cost on information processing costs and details how these costs can be mapped to those required as inputs to the framework obtained using the model described in Section \ref{System_Model}.

	\subsection{Multi-Layer System Model}\label{System_Model}
	This work, inspired by the model in \cite{FB'97} and extending upon our previous works (\hspace{-1mm} \cite{ICC'14} and \cite{WiOpt'14}), considers a network model consisting of four independent layers where each layer represents a particular network component. The four network components considered in this paper are users, base stations, backhaul nodes, and data centers. The lowest layer (layer $0$) consists of users\footnote{Please note the term user in this work denotes only the \textit{active} users in the network.} represented by a homogeneous Poisson process $\Phi_0 \subset \mathbb{R}^2$ with intensity $\lambda_0>0$. Similarly, the topmost layer (layer $3$) consists of data centers modeled by a homogeneous Poisson process $\Phi_3 \subset \mathbb{R}^2$ with intensity $\lambda_3>0$. Layer $2$ consists of both fiber optic and microwave backhaul nodes which are modeled using a stationary mixed Poisson process (see \cite{Stoyan'95} for details) $\Phi_2 \subset \mathbb{R}^2$ with a randomized intensity function $X$ having a two-point distribution 
\begin{equation*}
\mathbb{P}\left(X = \lambda_{2\text{MW}} \right) = p, \, \mathbb{P}\left(X = \lambda_{2\text{OF}} \right) = 1-p, 
\end{equation*}
where $0<p<1$ is the probability of having a microwave backhaul and $(1-p)$ is the probability of having a fiber optic backhaul. Hence, the intensity of $\Phi_2$ is 
\begin{equation*}
\lambda_2 = p\lambda_{2\text{MW}}+(1-p)\lambda_{2\text{OF}},
\end{equation*}
where $\lambda_{2\text{MW}}>0$ is the intensity of the microwave backhaul and $\lambda_{2\text{OF}}>0$ is the intensity of the fiber optic backhaul.
\begin{figure}[!b]
  \centering
  \includegraphics[width=0.9\linewidth]{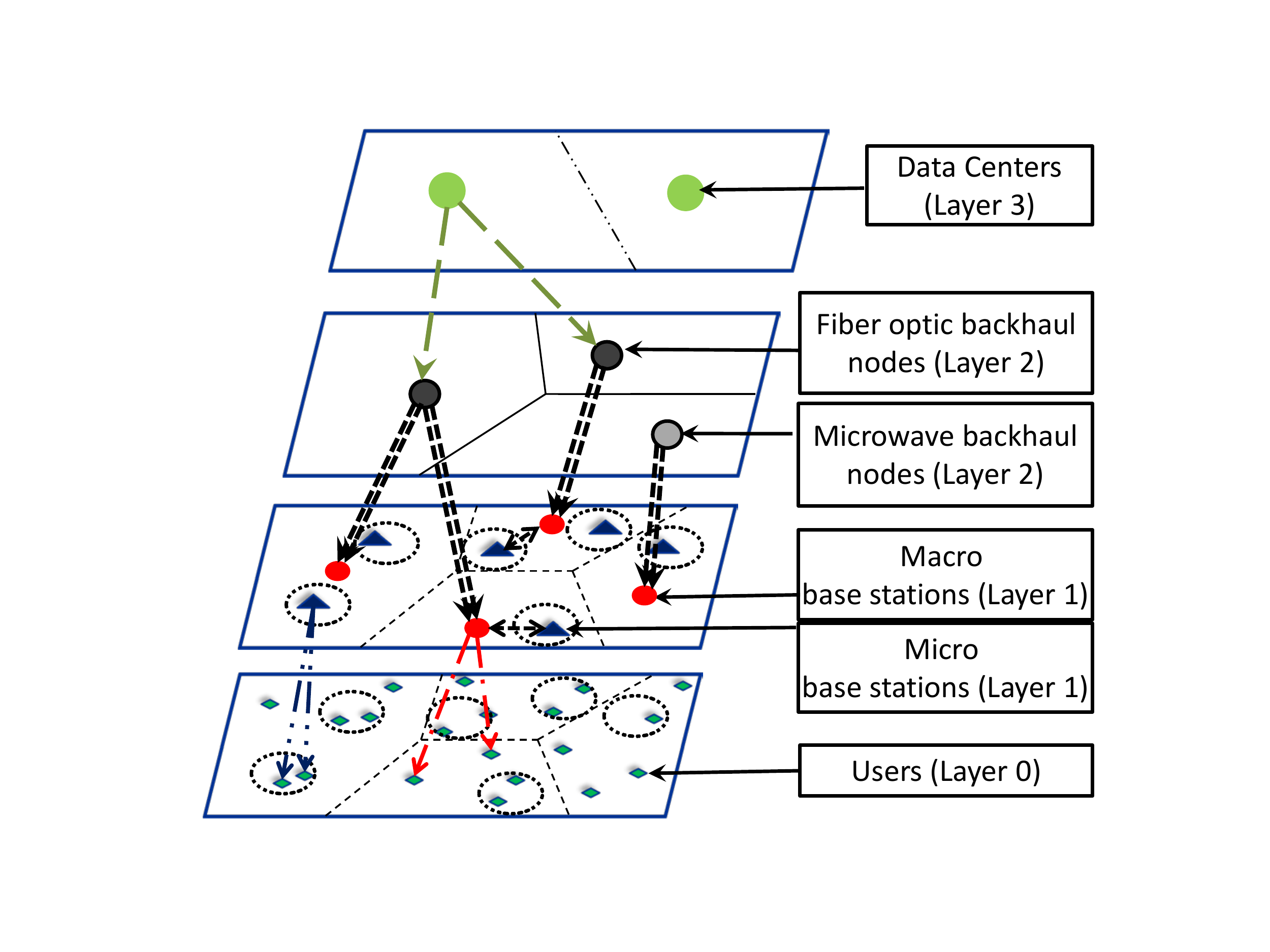}
  \caption{An illustration of a $4$ layer network model.}
  \label{fig:model}
  \end{figure}
The penultimate layer (layer $1$) consists of base stations (macro and micro base stations) represented by a stationary Poisson cluster process $\Phi_1 \subset \mathbb{R}^2$ consisting of two parts: cluster centers representing macro base stations and cluster members representing micro base stations. The cluster centers are modeled by a stationary Poisson point process $\Phi_{1c} \subset \mathbb{R}^2$ with intensity $\lambda_{1c}>0$, and conditioned on $\Phi_{1c}$, the cluster members are modeled by an inhomogeneous Poisson point process $\Phi_{1m} \subset \mathbb{R}^2$ with intensity function
\begin{equation*}
\rho(y) = \lambda_{1m}\sum\limits_{x \in \Phi_{1c}} f(y-x), \; y \in \mathbb{R}^2,
\end{equation*}
where $\lambda_{1m}>0$ is the expected number of cluster members around each cluster center and $f(\cdot)$ is a continuous density function which describes how a cluster member (micro base station) is distributed around a cluster center (macro base station). It is important to note that the cluster intensity and the normalized kernel bandwidth\footnote{Kernel bandwidth denotes the spread of the cluster points around the parent points and should not be confused with ``bandwidth" in communications.} are equal and fixed. Hence, $\Phi_{1m}$ is a shot noise Cox process (see \cite{Moeller'02}) and can also be considered as a Neyman-Scott cluster process (see \cite{NS'58}). This implies $\Phi_{1m}$, when not conditioned on $\Phi_{1c}$, is stationary with intensity $\lambda_{1c}\lambda_{1m}$. Thus, the superposition $\Phi_1 = \Phi_{1c} \cup \Phi_{1m}$ forms a stationary Poisson cluster process with intensity $\lambda_1=\lambda_{1c}(1+\lambda_{1m})$. The model applied in this paper is similar to \cite{Twireless'14} and different to \cite{Ganti'09}, which models ad-hoc networks.  The network can be visualized as shown in Fig. \ref{fig:model}. This paper  assumes that only connections between adjacent layers are allowed, for example, backhaul nodes cannot communicate with the users directly. This work also \textit{does not} explicitly consider costs incurred while connecting backhaul nodes to each other as well as rate improvements due to the use of Coordinate Multi-Point (CoMP) techniques \cite{3GPP.TR.36.819}. 

An important factor that determines the final deployment cost is the number of users that the network needs to cater to and the number of users that are connected to a network component, i.\,e., the number of users (or the number of points of $\Phi_0$) that are connected to a given point $x$ in layer $\Phi_i$ for $i\ge 1$. This is denoted by $\mathcal{N}_{x}$ and gives the total number of points in a subtree (as seen in Fig.\ref{fig:model}). The Voronoi tessellation (see \cite{Moeller'94}) determines which points of the lower layer are connected to a particular point in the upper layer. By this assumption, it is implicit that the users connect to their nearest base station. In general, the cell centered at a point $x$ belonging to process $\Phi_i$ is denoted by $V_x(\Phi_i)$. This structure is used in the following sections to estimate the cost of deploying a node in the backhaul layer. However, as shown in \cite{ICC'14}, defining cell areas for associating users with their respective base stations can also be based on a Signal-to-Interference-plus-Noise (SINR) tessellation which is different from the Voronoi tessellation. For more details about the SINR tessellation the reader is referred to works such as \cite{ICC'14}, \cite{FB'09}, and \cite{JA-FB'10}. 


	
	\subsection{Data Processing Model}\label{Complexity_Model}
	The analysis in this section relates costs induced by operating the communication infrastructure to costs required for processing information in a Cloud-\ac{RAN} system with a heterogeneous backhaul. This translation (or mapping) of costs is required to study the effect of information processing on the deployment cost of a network. The communication infrastructure spans multiple layers and is composed of the base station layer as well as the backhaul layer. Furthermore, the data center layer also processes all incoming and outgoing transmissions. The processing requires ample computational resources which are dependent directly on the parameterization and operating regime of the mobile network. A major contributor to the cost of a Cloud-\ac{RAN} system is the data center layer. If a data center is provided with too few computational resources, computational outage occurs. In which case, a transmission may not be successfully decoded though the channel quality is satisfactory. In contrast, if the system is over provisioned, it will be underutilized most of the time, which reduces the cost effectiveness of a centralized system. In general, the processing requirements can be separated into those for uplink and downlink. As shown in \cite{Bhaumik.Chandrabose.Jataprolu.Kumar.Muralidhar.Polakos.Srinivasan.Woo.MobiCom.2012} based on the OpenAir platform (see \cite{OpenAir.2014} for details), the processing requirements in the uplink exceed those of the downlink by a factor of $5$ or more. A majority of the uplink processing resources are required for the decoder, i.e., more than $\unit[80]{\%}$ of the overall expected uplink processing resources. Additionally, while the processing demand in the downlink is fairly predictable, it is highly variable in the uplink. For the sake of brevity, we assume symmetric uplink and downlink traffic in this paper. Furthermore, based on the results in \cite{Bhaumik.Chandrabose.Jataprolu.Kumar.Muralidhar.Polakos.Srinivasan.Woo.MobiCom.2012}, we assume that the processing requirements of the downlink as well as the higher layers of the uplink are about $\unit[40]{\%}$ of the expected processing resources of the uplink turbo decoder. It is important to note that the framework presented in this paper can also co-opt asymmetric traffic patterns and different processing distributions without changing the theoretic findings.

  \begin{figure}
    \centering
  \includegraphics[width=8cm, height=8cm, keepaspectratio, viewport = 132 357 473 513]{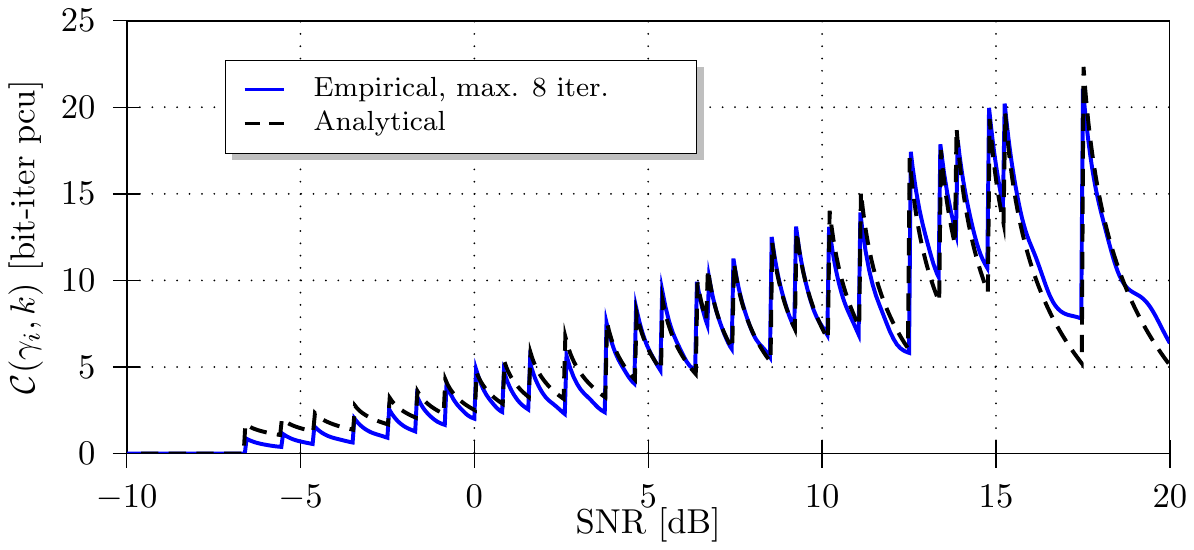}
	\vspace{-3mm}
    \caption{A comparison of theoretical and actual LTE decoder complexity \cite{Rost.Talarico.Valenti.TWC.2014}.}
    \vspace{-5mm}
    \label{fig:complexity_model:complexity_over_snr}
  \end{figure}
The uplink processing complexity is determined by the turbo decoder. Let $\gamma$ denote the channel's instantaneous \ac{SNR} and $k$ denote the index of the \ac{MCS} chosen. Then, the complexity of the decoding process scales both with the number of information bits $I(\gamma, k)$ processed and the number of iterations $L(\gamma, k)$ required to decode a codeword of duration $t_s$ and bandwidth $B_s$. In the following, we use the normalized data processing requirement $\mathcal{D}_\text{norm}(\gamma, k) = I(\gamma, k)L(\gamma, k)/(t_s B_s)$ to quantify the required data processing resources, i.e., 
$\mathcal{D}_\text{norm}$ provides a measure of the decoding complexity required for each bit transmitted per channel use -- which is equivalent to the total complexity of decoding a codeword normalized by the temporal and spectral resources occupied by it. The normalized complexity is measured in \emph{bit-iterations per channel use} (pcu) (which is equivalent to \emph{bit-iterations per second per Hertz}). The advantage of using this measure is its independence from the resources occupied by the system (similar to spectral efficiency).

In contrast, Bhaumik et al. (in \cite{Bhaumik.Chandrabose.Jataprolu.Kumar.Muralidhar.Polakos.Srinivasan.Woo.MobiCom.2012}) performed a quantitative assessment of the required processing \emph{time} in an uplink LTE system, i.e., for a specific \ac{MCS} $k$, the time required to process one codeword has been determined. Although this empirical study is very important and valuable, it does not permit making generic inferences about the stochastic behavior of the uplink decoding process as well as the centralization gain in a Cloud-\ac{RAN} system. In \cite{Rost.Talarico.Valenti.TWC.2014}, the authors derive an analytical model which allows evaluating the data processing requirements of a set of $N_\text{cloud}$ base stations. Fig. \ref{fig:complexity_model:complexity_over_snr} (taken from \cite{Rost.Talarico.Valenti.TWC.2014}) shows a comparison of empirical measurements and the analytical model. It is apparent that complexity clearly varies with the \ac{SNR}. In fact, if the data rate $R_k$ of a chosen \ac{MCS} $k$ is close to capacity, the processing resources required also increase super-linearly. This is due to the fact that the number of turbo-decoder iterations required increases super-linearly as the system operates close to capacity. Since the complexity per iteration also increases with the \ac{SNR} (due to an increase in the number of bits transmitted), the variation in the data processing requirement also increases in proportion to the \ac{SNR}.
In order to evaluate the extent of this dependence, we introduce the model derived in \cite{Rost.Talarico.Valenti.TWC.2014} briefly and use it to quantify the data processing resources required in a Cloud-\ac{RAN} system.

The data processing resources required depend on the following parameters which originate
from the decoder power consumption model in \cite{Grover.Woyach.Sahai.JSAC.2011} (and is applied in \cite{Rost.Talarico.Valenti.TWC.2014}). The first parameter chosen is the \ac{MCS} $k$ which determines the code rate $R_k$ at which the system is operated. Then, we consider a target channel outage probability $\hat\epsilon_\text{channel}$ and a complexity scaling function $K(\hat\epsilon_\text{channel})$. The decoder connectivity $\zeta$, which models the connectivity of the message passing algorithm, is defined and considered in \cite{Grover.Woyach.Sahai.JSAC.2011}. Based on an empirical study, \cite{Rost.Talarico.Valenti.TWC.2014} details a suitable parameterization for a 3GPP LTE decoder whose parameters are $\zeta = 6$ and $K(\hat\epsilon_\text{channel}) = 0.2$ with $\hat\epsilon_\text{channel} = 0.1$. Based on this model, the authors of \cite{Rost.Talarico.Valenti.TWC.2014} derive the functional dependence of these variables and show that the complexity of the decoder is well approximated by
\begin{align}
\mathcal{D}_\text{norm}(\gamma, k) \approx & \frac{R_k}{\log_2 \left(\zeta - 1\right)} \bigg(\log_2\left(\frac{\zeta - 2}{K(\hat\epsilon_\text{channel})\zeta}\right) - \nonumber \\[5pt]
& 2\log_2\left(\log_2(1 + \gamma) - R_k\right) \bigg).
\label{eq:complexity.model:100}
\end{align}
Throughout this paper, we assume that signals from each base station are processed independently, i.e., multi-cell signal processing is not considered. Multi-cell signal processing has the potential to increase the system throughput, but it also requires additional data processing resources. Since both multi-cell signal processing and data processing resources are highly dependent on the underlying channel and traffic assumptions (and the exact nature of their relationship is hard to determine), we, therefore, do not consider multi-cell signal processing in this work. In this paper, we compare a centralized \ac{RAN} implementation and a distributed implementation, where base station signals are processed locally at each base station (referred to as \ac{DRAN}). A summary of the differences between Cloud-\ac{RAN} and \ac{DRAN} is given in Table \ref{tab:CloudRAN_vs_DRAN}. In the case of centralized implementation, we assume that forward error correction and protocol layer functionality described above is performed at the central processing entity using high-volume IT hardware \cite{Wuebben.etal.SPM.2014}.
\begin{table*}[t]
  \centering
  \caption{Summary of the differences between Cloud-\ac{RAN} and \ac{DRAN}}
  \label{tab:CloudRAN_vs_DRAN}
  \begin{tabular}{|l||p{7cm}|p{7cm}|}
  \hline
   ~ & \multicolumn{1}{c}{Cloud-\ac{RAN}} & \multicolumn{1}{|c|}{\ac{DRAN}} \\ \hline\hline
    Base stations & Lower complexity and conceivably cheaper & Full implementation (standard complexity and cost) \\ \hline
    Diversity gains & Multiplexing and computational (multi-user) diversity gains & Only per-base station diversity gains \\ \hline
    Data processing & High-volume commodity hardware & Dedicated DSP and ASIC implementations \\ \hline
    Backhauling & Higher throughput required and the latency is in the order of few milliseconds & Lower overhead \\ \hline
    Flexibility & Driven by software & Driven by hardware \\ \hline
    Programmability & Based on GPP & Based on DSP \\ \hline
  \end{tabular}
\end{table*}

Another critical aspect of system parameterization is the link-adaptation, which determines the minimum \ac{SNR} for which a specific \ac{MCS} $k$ is chosen. Ideally, if perfect random codes of infinite length are assumed, we can use the \ac{SNR} thresholds $\gamma^c_{k} = 2^{R_k} - 1$, which denote the threshold for rate $R_k$ in the Shannon capacity case. However, a practical system does not operate at capacity and the \ac{SNR} thresholds of a decoder (which is deployed in practice) are given by $\gamma^r_{k} =\nu \, \gamma_{\text{offset}} \,\gamma^c_{k}$ where $\nu = \unit[0.2]{dB}$ is a complexity calibration parameter for a system with a maximum of eight turbo-iterations. Furthermore, $\gamma_{\text{offset}}$ is an additional link-adaptation offset which allows the system complexity to be further reduced. The smaller the value of $\gamma_{\text{offset}}$, the closer the system operates to capacity but greater the number of iterations necessary to decode a codeword. If the offset is higher, fewer processing resources are required which implies the existence of an inherent trade-off between processing resources provided (and their associated costs) and the performance delivered.

  \begin{figure}[!t]
    \centering
    \subfigure[Outage processing resources per base station]{\includegraphics[width=7cm, height=7cm, keepaspectratio, viewport = 131 348 360 526]{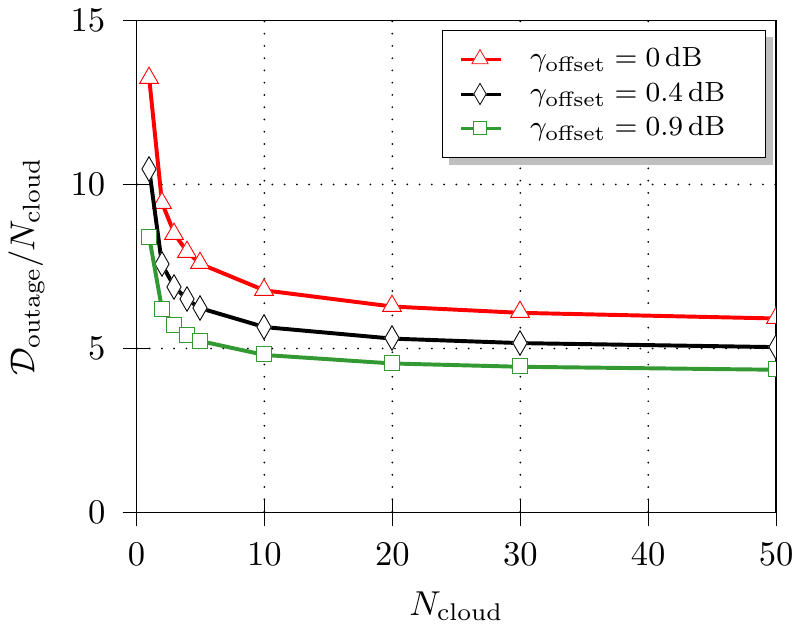}}
    \subfigure[Processing demand in number of servers]{\includegraphics[width=7cm, height=7cm, keepaspectratio, viewport = 132 348 360 526]{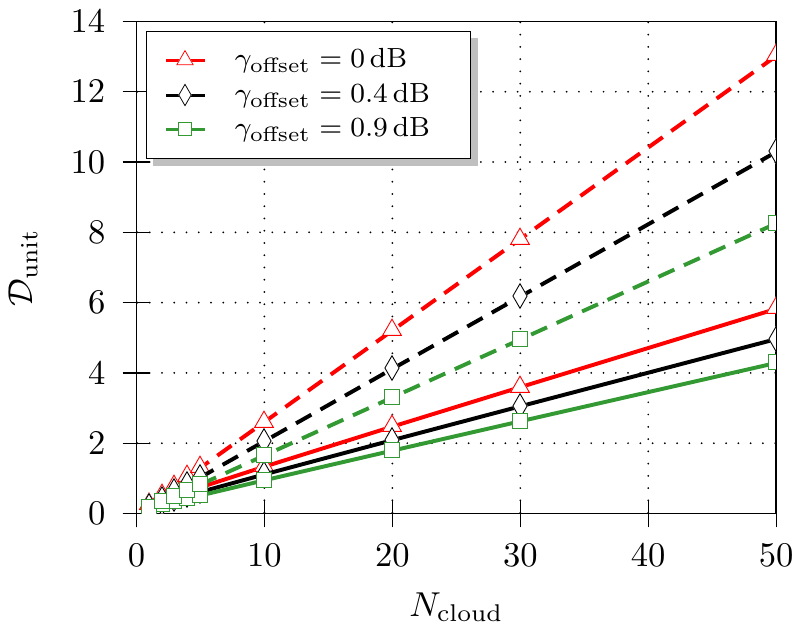}}
    \caption{Data processing capabilities required to ensure per-base-station computational outage probability $\hat\epsilon_\text{comp}$ \cite{Rost.Talarico.Valenti.TWC.2014}. Dashed lines denote the equivalent processing power required in a distributed implementation, i.e., \ac{DRAN}.}
    \vspace{-5mm}
    \label{fig:complexity_model:complexity_network}
  \end{figure}
Based on this model, \cite{Rost.Talarico.Valenti.TWC.2014} derives expressions for the mean $\Exp\left[\mathcal{D}_\text{norm}(\gamma, k)\right]$ and the variance $\text{Var}\left[\mathcal{D}_\text{norm}(\gamma, k)\right]$ of the processing resources. Other critical system parameters are the processing outage probability $\epsilon_\text{comp}(N_\text{cloud})$ and the outage processing demand $\mathcal{D}_\text{outage}(\hat\epsilon_\text{comp}, N_\text{cloud})$, which are related by
\begin{align}
   \epsilon_\text{comp}(N_\text{cloud}) & = 
   \Prob\left[
       \sum_{i=1}^{N_\text{cloud}} \mathcal{D}_\text{norm}(\gamma_i | k) > \mathcal{D}_\text{outage}(\hat\epsilon_\text{comp}, N_\text{cloud})
   \right]. 
  \nonumber 
\end{align}
$\mathcal{D}_\text{outage}$ quantifies the data processing resources required to guarantee a maximum per-base-station computational outage probability $\hat\epsilon_\text{comp} = 0.1$. Fig. \ref{fig:complexity_model:complexity_network}(a) shows the normalized outage resources as a function of the number of centralized base stations $N_\text{cloud}$ and the decoder quality characterized by $\gamma_{\text{offset}}$.

Using Fig. \ref{fig:complexity_model:complexity_network}(a), we can estimate the required normalized data processing complexity in \emph{bit-iterations per channel use} ($\mathcal{D}_\text{outage}$). We consider a 3GPP LTE system with $\unit[10]{MHz}$ bandwidth where one user may occupy at most $45$ physical resource blocks, each consisting of $7$ symbols $\times 12$ sub-carriers in a sub-frame of duration $t_s = \unit[0.5]{ms}$. Therefore, the data processing resources required are given by $\mathcal{D}_\text{abs} = \mathcal{D}_\text{outage} \times 45 \times 12 \times 7/\unit[0.5]{ms} \approx \mathcal{D}_\text{outage} \times 7.5 \times 10^6 \, \nicefrac{cu}{s}$ $[\nicefrac{\text{bit-iter}}{s}]$. A typical turbo-decoder implementation requires up to $1,000$ \acp{FLOP} per bit-iteration (see \cite{Valenti.Sun.IJWIN.2001}), i.e., the data processing demand required can be determined by $\mathcal{D}_\text{flops}=\mathcal{D}_\text{abs}\times 1000 \, \nicefrac{\text{FLOP}}{\text{bit-iter}}$ $[\nicefrac{\text{FLOP}}{s}]$. As a reference, the Intel Xeon 4870 is a 10 core processor which achieves $96 \, \nicefrac{\text{GFLOP}}{\text{s}}$ and is typically packaged on a quad socket board. Such a server setup (including 128 GB RAM) would cost about $\$ \,20,000.00$. If $\mathcal{D}_\text{unit} = \mathcal{D}_\text{flops} / (4\times 96 \, \nicefrac{\text{GFLOP}}{\text{s}})$ is the data processing demand in a number of such server setups (each server equipped with four processors), then Fig. \ref{fig:complexity_model:complexity_network}(b) shows the processing power required to operate a Cloud-\ac{RAN} system for three different offset values of the link-adaptation process. The dashed lines in Fig \ref{fig:complexity_model:complexity_network}(b) show the equivalent processing resources in \ac{DRAN}. These values will be used in Section~\ref{Obtaining_Cost_Values}  and Section \ref{Numerical_Evaluation}  to compute the network deployment cost.

\section{Cost Model}\label{Cost_Model}

	\subsection{Components of Deployment Cost}\label{Cost_Description}
	Typical deployment costs incurred by a service provider can broadly be classified into equipment cost, capacity cost, and infrastructure cost. The \textit{equipment cost} $C_i$ (in \$/device) represents the cost of a device deployed in a particular layer $i$. We assume that users buy their handset, and hence, the cost (to the service provider) $C_0=0$. The equipment cost $C_1$ is the cost of deploying a typical base station cluster consisting of one macro base station and $\lambda_{1m}$ micro base stations (on average). The equipment cost of a backhaul node is $C_2$ which is a linear combination of $C_{2\text{MW}}$ and $C_{2\text{OF}}$, where $C_{2\text{MW}}$ and $C_{2\text{OF}}$ are the equipment costs of a microwave backhaul device and a fiber optic backhaul device, respectively. The equipment costs of base stations and backhaul nodes can be written as 
\begin{align*}
C_1 &= \frac{C_{\text{macro}} + \lambda_{1m}C_{\text{micro}}}{(1+\lambda_{1m})}, \\
C_2 &= p\lambda_{2\text{MW}} C_{\text{MW}} + (1-p)\lambda_{2\text{OF}}C_{\text{OF}}, 
\end{align*}
where $C_{\text{macro}}$ and $C_{\text{micro}}$ are the equipment costs of a macro base station and a micro base station, respectively. Note that the cost $C_1$ is the cost of deploying a single cluster consisting of one parent (macro base station) and an average of $\lambda_{1m}$ cluster members (micro base stations). Hence, the equipment cost of the (entire) base station layer is 
\begin{equation*}
\lambda_1 C_1 = \lambda_{1c}C_{\text{macro}} + \lambda_{1c}\lambda_{1m}C_{\text{micro}}.
\end{equation*}
Finally, the equipment cost of a data center is $C_3$. It is important to note that $C_3$ is non-zero only in cases where the Cloud-\ac{RAN} is implemented. In the \ac{DRAN} case, $C_3 = 0$.

Since all point processes in our model are stationary, for mathematical simplicity, a point under consideration in the higher layer is assumed to be at the origin $o$. Then, the \textit{capacity cost} is the cost of connecting a device at point $x$ in layer $i$ to another device at point $o$ in layer $i+1$ for a given capacity requirement. This cost is considered to be of the form $A_{i,i+1} \, g(\|x\|)$, where $A_{i,i+1}>0$ is the base cost to be able to provide a certain capacity (or data rate) and $g(\|x\|)$ is a function of the distance $\|x\|$ which determines how the base cost scales with distance. For simplicity and to include all possible rates of polynomial increases in cost, $g(\|x\|)$ can be considered to be in power law form given by $g(\|x\|) = \|x\|^{\beta_{i,i+1}}$ where $\|x\|$ is the distance between the points $o \in \Phi_{i+1}$ and $x \in \Phi_i$, and $\beta_{i,i+1} \ge 0$ is the exponent based on which the cost increases. E.g., if $\beta_{i,i+1} = 1$, $A_{i,i+1} \, g(\|x\|) = A_{i,i+1} \, \|x\|$ which implies that the capacity cost increases linearly with distance. It is important to note that the base capacity cost $A_{i,i+1}$ can consist of many different components such as connectivity cost, etc. In this work, we assume the base capacity costs $A_{0,1}$ and $A_{1,2}$ to be fixed\footnote{These values are stated in Section \ref{Obtaining_Cost_Values} along with their sources. Note that this implies that the cost of connecting a user to a base station is dependent on the user demand and not on the type of base station catering to the demands.} and focus on the base capacity cost of data centers, i.e., $A_{2,3}$, since our intention is to find the dependence of deployment costs on processing and communication costs in cloud based radio access networks. The base capacity cost of data centers is dependent on two main factors. We call the first as ``\textit{capacity delivery cost}'', $A'_{2,3}$, which is the cost of delivering a given data rate to a particular distance. The second is the ``\textit{data processing cost}'', $A''_{2,3}$, which is dependent on the number of users and their demands, but does not scale with distance, i.e., $\beta_{2,3} = 0$ for $A''_{2,3}$. The complexity model detailed in Section \ref{Complexity_Model} forms the basis for defining the data processing cost between the backhaul and the data centers, which will be elaborated upon in Section~\ref{Obtaining_Cost_Values}. Therefore, the base capacity cost of a data center can be written as $A_{2,3} = A'_{2,3} + A''_{2,3}$. 

Likewise, \textit{infrastructure cost} is defined as the expense incurred to ensure that a point $x$ of layer $i$ and the point $o$ in layer $i+1$ are connected. It is considered to be of the form $B_{i,i+1} \, h(\|x\|)$, where $B_{i,i+1}>0$ is a quantity similar to $A_{i,i+1}$ (defined as the base cost for a particular type of installation) and $h(\|x\|)$ is a function of the distance $\|x\|$ between the two points under consideration. Once again, for ease of computation, $h(\|x\|)$ is taken to be $\|x\|^{\theta_{i,i+1}}$ where $\theta_{i,i+1} \ge 0$ determines how fast the base infrastructure cost increases with distance. Though the definitions of capacity and infrastructure costs are similar, the reason for considering them separately is as follows. Infrastructure cost is the cost incurred while laying the cable or installing microwave equipment which increases with the distance between the two points to be connected (due to labor charges, etc.). Furthermore, each technology has the ability to deliver a given data rate (with minimal losses) up to a particular distance for a fixed cost. Now, if the data rate desired is more than what a single installation of a particular technology can provide, it requires more than one installation of the same technology. Since additional installations can use the same physical route (e.g., cabling along the same path) as the initial installation, there is no (or negligible) infrastructure cost but there is added expenditure to meet capacity requirements, such as upgrading certain components, spectrum costs (in the case of a microwave backhaul), etc. Hence, the need for a separate cost category which we term as the capacity cost.

	\subsection{Computing the Deployment Cost}\label{Cost_Framework}
	If $\mathcal{C}_{\Phi_3}$ is the expected cost of deploying a device in the data center layer. Then, we get Theorem \ref{Th:Theorem1} whose proof as well as the expression for the Palm distribution $\mathbb{P}_1^o\left( \cdot \right)$ are given in Appendix~\ref{Theorem_Proof}. Therefore, the total cost of such a network is given by 
\begin{equation}
	\label{eq:total_cost}
	C_{\text{TOT}} =  \lambda_3\left(C_3 + \mathcal{C}_{\Phi_3}\right).
\end{equation} 
Note that though expressions for $\Psi_3(\cdot)$ and $\Psi_4(\cdot)$ are not closed form expressions like those obtained for the other terms, solving them numerically is quite simple and takes only a few seconds on commercially available software.
\newtheorem{theorem}{Theorem}
\begin{figure*}[t]
\centering
\begin{theorem}
\label{Th:Theorem1}
In a $4$-layer model that uses power law functions to describe capacity and infrastructure costs, the expected cost of deploying a data center is given by 
\begin{align}
\label{eq:theorem1}
\mathcal{C}_{\Phi_3} &= \frac{\lambda_{2}}{\lambda_3} \, \Bigg[C_{2} \, + \frac{\lambda_0}{\lambda_2} \bigg(A'_{2,3} \frac{\Gamma\left(\frac{\beta_{2,3}}{2} + 1 \right)}{\left(\pi \lambda_3 \right)^{\beta_{2,3}/2}} + A''_{2,3} \bigg) + B_{2,3} \frac{\Gamma\left(\frac{\theta_{2,3}}{2} + 1 \right)}{\left(\pi \lambda_3 \right)^{\theta_{2,3}/2}} + \frac{\lambda_1}{\lambda_2} \, \Bigg\{ \, C_1 \, + 
\frac{\lambda_0}{\lambda_1}\Psi_1\big(A_{1,2}, \beta_{1,2}, \lambda_{2\text{MW}}, \lambda_{2\text{OF}}, p \big) \, +  
  \nonumber \\[5pt] 
& \Psi_2\big(B_{1,2}, \theta_{1,2}, \lambda_{2\text{MW}}, \lambda_{2\text{OF}}, p \big) \, + 
\frac{\lambda_0}{\lambda_1} \Big[  \Psi_3\big(A_{0,1}, \beta_{0,1}, \lambda_{1c}, \lambda_{1m}, f(\cdot), \sigma^2 \big) + \Psi_4\big(B_{0,1}, \theta_{0,1}, \lambda_{1c}, \lambda_{1m}, f(\cdot), \sigma^2 \big) \Big] \Bigg\} \Bigg]
\end{align}
where 
\begin{gather*}
\Psi_1\big(A_{1,2}, \beta_{1,2}, \lambda_{2\text{MW}}, \lambda_{2\text{OF}}, p \big) = A_{1,2} \Bigg[ \frac{p \, \Gamma(\frac{\beta_{1,2}}{2}+1)}{(\pi\lambda_{2\text{MW}})^{\beta_{1,2}/2}} + \frac{(1-p) \, \Gamma(\frac{\beta_{1,2}}{2}+1)}{(\pi\lambda_{2\text{OF}})^{\beta_{1,2}/2}} \Bigg],\\
\Psi_2\big(B_{1,2}, \theta_{1,2}, \lambda_{2\text{MW}}, \lambda_{2\text{OF}}, p \big) = B_{1,2} \Bigg[ \frac{p \, \Gamma(\frac{\theta_{1,2}}{2}+1)}{(\pi\lambda_{2\text{MW}})^{\theta_{1,2}/2}} + \frac{(1-p) \, \Gamma(\frac{\theta_{1,2}}{2}+1)}{(\pi\lambda_{2\text{OF}})^{\theta_{1,2}/2}} \Bigg],\\
\Psi_3\big(A_{0,1}, \beta_{0,1}, \lambda_{1c}, \lambda_{1m}, f(\cdot), \sigma^2 \big) = A_{0,1} \, \lambda_1 \int\limits_{\mathbb{R}^+} r^{\beta_{0,1}} \mathbb{P}_1^o\left( b(o,r) \right) \mathrm{d}r,\\
\Psi_4\big(B_{0,1}, \theta_{0,1}, \lambda_{1c}, \lambda_{1m}, f(\cdot), \sigma^2 \big) = B_{0,1} \, \lambda_1 \int\limits_{\mathbb{R}^+} r^{\theta_{0,1}} \mathbb{P}_1^o\left( b(o,r) \right) \mathrm{d}r,
\end{gather*}
wherein $\mathbb{P}_1^o\left( \cdot \right)$ is the Palm distribution with respect to $\Phi_1$.
\end{theorem}
\hrulefill
\end{figure*}



	\subsection{Obtaining Values for Cost Calculation}\label{Obtaining_Cost_Values}
	From equation (\ref{eq:theorem1}), in order to observe the impact of processing and communication costs on the deployment cost, we gather that methods to determine the appropriate values of base costs as well as intensities of various network components are essential. This subsection details how the intensities of the various devices in the network as well as the base costs for the respective devices are chosen. It is important to note that these values serve a purely illustrative purpose and the accuracy of the values assumed is not the primary focus of this work.

The user intensity is assumed to be $\lambda_0 = 170/\text{km}^2$ and the average user demand is assumed to be $10$ Mbps based on the FTP model in \cite{3GPP.TR.36.814}. Next, we use a result from our previous work, \cite{VTC'12}, to find the base station intensity. The expression
\begin{align}
\label{eq:SAR}
& \bar{R}_{\Phi_1}\left(\lambda_1,\lambda_0, P_{\text{Tx}}, \sigma_{\text{N}}^2\right) = \nonumber \\[5pt] 
& \frac{\pi^{5/2}}{2}\sqrt{\frac{\lambda_0\lambda_1 P_{\text{Tx}}}{\sigma_{\text{N}}^2}} \, \text{Erfc}\left[\frac{\pi^2\lambda_0}{4}\sqrt{\frac{P_{\text{Tx}}}{\sigma_{\text{N}}^2}}\right] \exp{\left[\frac{\pi^4\lambda_0^2P_{\text{Tx}}}{16\sigma_{\text{N}}^2}\right]}, 
\end{align}
provides a relationship between the spectral efficiency ($\bar{R}_{\Phi_1}(\cdot)$), user intensity, base station intensity, transmit power ($P_{\text{Tx}}$), and noise power ($\sigma_{\text{N}}^2$). It must be noted that though \cite{VTC'12} models base stations as a homogeneous Poisson point process and this work considers a stationary Poisson cluster process to model them, the following reasons make it viable to use equation (\ref{eq:SAR}). Firstly, computing the base station intensity from the expression for spatially averaged rate in a network modeled using a Poisson cluster process is extremely complicated and laborious (see \cite{Twireless'14} for more details). Secondly, since we are interested only in the total number of base stations in an area for computing deployment costs, using equation (\ref{eq:SAR}) provides a much simpler alternative. Moreover, this paper assumes a strict Voronoi tessellation (implying that the entire area is covered) and hence, the only aspect of importance (in our scenario) is whether or not the average user demands are met. Under these assumptions, finding the ``total'' base station intensity required to satisfy average user demands should suffice. Lastly, the model in \cite{VTC'12}, which is used to derive equation (\ref{eq:SAR}), considers a fixed transmit power per user\footnote{E.g., if 5 users are connected to a base station then its total transmit power would be $5P_{\text{Tx}}$. Hence, this model is similar to a Time Division Multiple Access (TDMA) system.} due to which a single value for the transmit power of both macro base stations and micro base stations can be assumed. 
However, note that the \textit{total} transmit powers of the macro and micro base stations are different since macro base stations have larger coverage areas owing to which they serve a greater number of users. 

Now, the transmit power (in dBm) is given as $P_{\text{Tx}} = 18.22 + 10\log_{10}{\left(\text{No. of sub-carriers}\right)} + 30$ and the noise power (in dBm) is calculated by $\sigma_{\text{N}}^2 = -174 + 10 \log_{10}{\left(B\right)}$ for a bandwidth $B$. Since we consider an LTE system with 10 MHz bandwidth, $29\%$ control overhead, $600$ sub-carriers, and $15 $ KHz sub-carrier spacing, we get $P_{\text{Tx}}=46$ dBm and $\sigma_{\text{N}}^2 = -146.22$ dBm. In order to ensure user satisfaction, the service provider has to provide a spectral efficiency that is at least equal to what the users expect. As before, assuming the average user demand to be $10$ Mbps, results in a spectral efficiency requirement of $1.0847$ bps/Hz. Substituting the values of user intensity, transmit power and noise power (in Watts), and spectral efficiency in equation (\ref{eq:SAR}) results in a base station intensity\footnote{Though the value of $\lambda_1$ is rather high and it translates to an inter-site distance of approximately $90$m (if a hexagonal grid deployment is assumed), this value is in keeping with the consensus that heterogeneous networks (of the future) could have inter-site distances of about $100$m or less.} $\lambda_1 = 50.03/\text{km}^2$ for the \ac{DRAN} architecture, which is assumed to be the baseline for comparison. However, as seen in Section \ref{Complexity_Model}, the spectral efficiency that can be achieved varies depending on the quality of the decoder and is represented by the link-adaptation offset $\gamma_{\text{offset}}$. Hence, if $\gamma_\text{offset} > \unit[0]{dB}$, we have to account for a rate offset $\Delta R(\gamma_\text{offset})$ which must be compensated for by a corresponding change in the base station intensity $\lambda_1$. Based on the results  obtained in \cite{Rost.Talarico.Valenti.TWC.2014} after normalization, we get $\Delta R(\unit[0.4]{dB})=\unit[0.01322]{bps/Hz}$ and $\Delta R(\unit[0.9]{dB})=\unit[0.029751]{bps/Hz}$. Hence, the spectral efficiency for each of these cases with non-zero offset is given by $1.0847 + \Delta R(\gamma_\text{offset})$. Then, using equation (\ref{eq:SAR}), with all other values remaining unchanged, results in base station intensities of $\lambda_1 = 51.2/\text{km}^2$ and $\lambda_1 = 52.8/\text{km}^2$ for rate offsets $\Delta R(\unit[0.4]{dB})$ and $\Delta R(\unit[0.9]{dB})$, respectively.

Then, for each value of $\lambda_1$, the individual intensities of macro and micro base stations can be chosen from the set of solutions to the equation $\lambda_{1c}(1+\lambda_{1m}) = \lambda_1$. Furthermore, assume the cluster variance\footnote{Recall that the cluster variance determines the spread of the micro base stations around a macro base station.} $\sigma^2 = 0.5$. This cluster variance ensures that the micro base stations are fairly widely scattered within the macro base station's cell area. Additionally, assume that it is equally likely that a particular transmission uses either a microwave backhaul or a fiber optic backhaul. This implies that $p = 0.5$ and $\lambda_2 = 0.5 \, \lambda_{2\text{MW}}+ 0.5 \, \lambda_{2\text{OF}}$. From information provided by a large European service provider, we gather that (on average) one fiber optic backhaul node or two microwave backhaul nodes are considered for about $15$ -- $20$ base stations. Therefore, we assume $\lambda_2 = 5/\text{km}^2$. Since the goal of this work is to observe the effects of information processing on the deployment costs, we do not assume any values for data center intensities. Section~\ref{Numerical_Evaluation} observes the changes in deployment cost when these intensities are varied. 

Finally, we detail the base costs of various devices and the rate at which they scale with the distance between two different devices. These costs (taken from \cite{Exalt},\cite{FOA}, and \cite{PIMRC'04}) are listed in Table \ref{tab:EquipCost} and Table \ref{tab:BaseCost}. The costs in Table \ref{tab:EquipCost} vary depending on the scenario considered, i.e., the \ac{DRAN} case with $\gamma_\text{offset}=\unit[0]{dB}$ or CRAN corresponding to Cloud-RAN. It is important to note that the references mentioned above contain a wide range of cost values for each device and the mean of the range (for each of these values) is considered in this paper. The data processing costs $A''_{2,3}$ are obtained using the slopes of the curves in Fig. \ref{fig:complexity_model:complexity_network}(b) and are tabulated in Table \ref{tab:BaseCapacityCost_A23}. They are used in Section \ref{Numerical_Evaluation} for the numerical evaluation.

\begin{table}
        \caption{Equipment Costs}
        \label{tab:EquipCost}
        \centering
       \begin{tabular}{|c|c|c|}
        \hline
        & \multicolumn{2}{c|}{Value (in \$)} \\ \cline{2-3}
        Type of Cost & DRAN & CRAN \\ \hline
        $C_{\text{macro}}$ & 50000 & 25000 \\ \hline
        $C_{\text{micro}}$ & 20000 & 10000 \\ \hline
        $C_{\text{MW}}$ & 50000 & 50000 \\ \hline
        $C_{\text{OF}}$ & 5000 & 5000 \\ \hline
        $C_3$ & 0 & 40000 \\ \hline
        \end{tabular}
\end{table}
\begin{table}
          \caption{Base Costs}
          \label{tab:BaseCost}
          \centering
          \begin{tabular}{|c|c|c|c|}
          \hline
          & & \multicolumn{2}{c|}{Type of Back-haul} \\ \cline{3-4}
          Type of Cost (in \$) & Values & Microwave & Optic Fiber \\ \hline
          \multirow{3}{*}{Capacity Cost} & $A_{0,1}$ & 5000 & 5000 \\ \cline{2-4}
          & $A_{1,2}$ & 5000 & 5000 \\ \cline{2-4} & $A'_{2,3}$ & 5000 & 5000 \\ \hline
          \multirow{3}{*}{Infrastructure Cost} & $B_{0,1}$ & 10000 & 10000 \\ \cline{2-4}
          & $B_{1,2}$ & 5000 & 100000 \\ \cline{2-4} & $B_{2,3}$ & 10000 & 100000 \\ \hline
          \end{tabular}
\end{table}
\begin{table}
 \caption{Data Processing Cost $A''_{2,3}$}
 \label{tab:BaseCapacityCost_A23}
 \centering
 \begin{tabular}{|c|c|c|}
 \hline
 $\gamma_\text{offset}$ & $\lambda_1$ & $A''_{2,3}$ (in \$) \\ \hline
 $0$ dB & $50.0/\text{km}^2$  & $(0.111 \!\cdot\! 50.0 + 0.0051) \!\cdot\! 20000 / 170 = \textbf{653.54}$ \\ \hline
 $0.4$ dB & $51.2/\text{km}^2$  & $(0.096 \!\cdot\! 51.2 + 0.0036) \!\cdot\! 20000 / 170 = \textbf{578.68}$ \\ \hline
 $0.9$ dB & $52.8/\text{km}^2$  & $(0.083 \!\cdot\! 52.8 + 0.0027) \!\cdot\! 20000 / 170 = \textbf{515.89}$ \\ \hline
 \end{tabular}
 \vspace{3mm}
  \caption{Exponents}
  \label{tab:Exponents}
  \centering
  \begin{tabular}{|c|c|c|}
  \hline
  & \multicolumn{2}{c|}{Type of backhaul} \\ \cline{2-3}
  Exponents & Microwave & Optic Fiber \\ \hline
  $\beta_{0,1}$ & 4 & 4 \\ \hline 
  $\beta_{1,2}$ & 2 & 1 \\ \hline
  $\beta_{2,3}$ & 2 & 1 \\ \hline
  $\theta_{0,1}$ & 2 & 2 \\ \hline
  $\theta_{1,2}$ & 2 & 1 \\ \hline
  $\theta_{2,3}$ & 2 & 1 \\ \hline
  \end{tabular}
\end{table}

The sources mentioned above, i.e. \cite{Exalt},\cite{FOA}, and \cite{PIMRC'04}, also provide the cost of a particular device and the range (in terms of radial distance) it can cover from which, the values of the exponents listed in Table \ref{tab:Exponents} have been extrapolated. The reasons for the choice of values in Table \ref{tab:Exponents} are as follows. The capacity cost between a user and base station is assumed to scale with distance according to the pathloss exponent. Consider a dense urban scenario which implies that the pathloss is approximately 4 (see \cite{Auer'11}), i.\,e., $\beta_{0,1}=4$. The data from \cite{FOA} indicates that the capacity cost for a fiber optic backhaul scales linearly with distance, i.e., farther the distance a given capacity has to be provided to, the greater would be the cost, i.\,e., $\beta_{1,2}=1$. Based on information from a large European operator, we assume $\beta_{2,3}= 2$  for a microwave backhaul which implies that the base capacity cost scales quadratically with distance. Similarly, with input from the same European operator, the base capacity cost for a fiber optic backhaul is assumed to scale linearly with distance, i.\,e., $\beta_{2,3}= 1$. For a microwave backhaul, extrapolating from data collected in \cite{Exalt}, we find that the capacity cost between a base station and a backhaul node scales approximately quadratically with distance, i.\,e., $\beta_{1,2}=2$. From the costs of base stations and their respective coverage areas\footnote{The costs in the reference are given based on the ability of a particular device to reach a particular radial distance.} in \cite{PIMRC'04}, we infer that the infrastructure cost between a user and a base station scales quadratically with distance, i.\,e., $\theta_{0,1}=2$. Based on data from \cite{FOA}, we assume that the infrastructure cost for a fiber optic backhaul scales linearly with distance. Hence, $\theta_{1,2} = \theta_{2,3}= 1$. Similar to the capacity cost for a microwave backhaul, it can be concluded that infrastructure cost increases quadratically with distance from \cite{Exalt}, i.\,e., $\theta_{1,2}=2$. Finally, the base infrastructure cost, based on information from operators, is assumed to scale quadratically with distance if a microwave backhaul is used, i.\,e., $\theta_{2,3}= 2 $.

It is important to reiterate that these values serve an illustrative purpose. Since the cost effectiveness documented in Section \ref{Numerical_Evaluation} is noted using the deployment costs calculated using the same values, the differences observed remain unchanged.

\section{Numerical Evaluation}\label{Numerical_Evaluation}
\begin{figure}
  \centering
    \subfigure[Cost (in millions of dollars) versus data center intensity for fiber optic links between backhaul nodes and data centers.]{\includegraphics[width=7cm, height=7cm, keepaspectratio, viewport = 131 348 360 526]{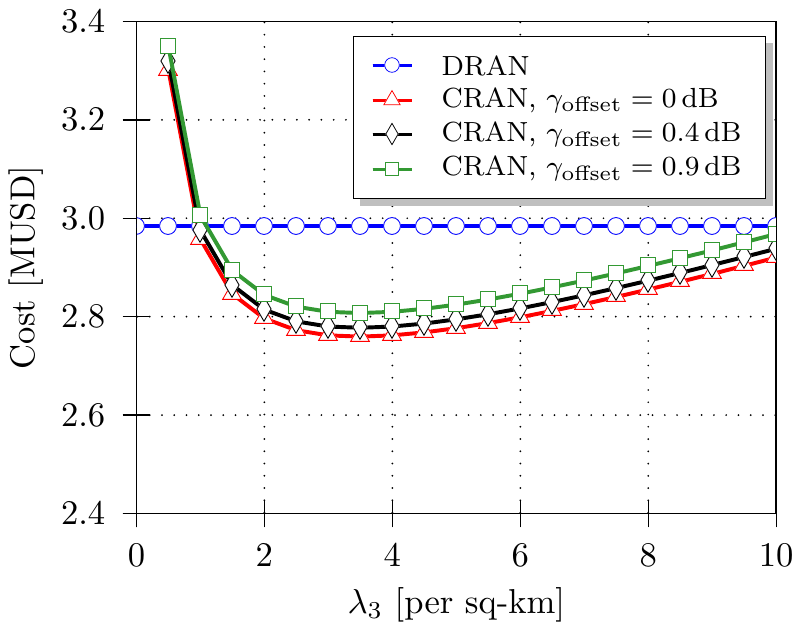} \label{Fig:Results.Cost.over.DC.and.RRHcost_a} }
    \subfigure[Cost (in millions of dollars) versus a decreasing difference between base station equipment costs in Cloud-\ac{RAN} and \ac{DRAN} for fiber optic links between backhaul nodes and data centers.]{\includegraphics[width=7cm, height=7cm, keepaspectratio, viewport = 131 348 360 526]{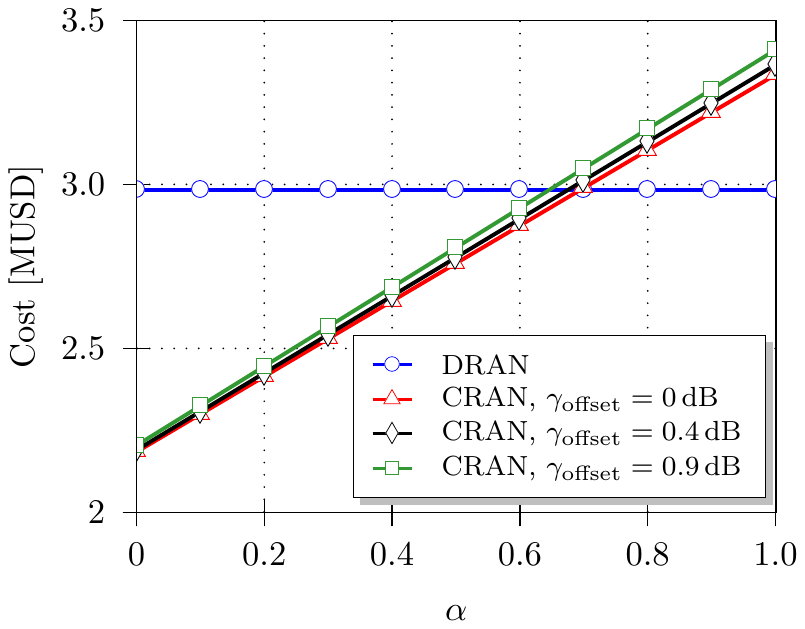} \label{Fig:Results.Cost.over.DC.and.RRHcost_b} }
\caption{Illustrating the variation in cost with increasing data center intensity and a decrease in the gap between base station equipment costs in Cloud-\ac{RAN} and \ac{DRAN} ($\alpha$). Note that all other parameters are fixed.}
\vspace{-5mm}
\label{Fig:Results.Cost.over.DC.and.RRHcost}
\end{figure}

The values described in Section \ref{Obtaining_Cost_Values} (and tabulated in Tables \ref{tab:EquipCost}, \ref{tab:BaseCost}, \ref{tab:BaseCapacityCost_A23}, and \ref{tab:Exponents}) are substituted in equation (\ref{eq:total_cost}) for the observations made in this section. Recall that for the \ac{DRAN} case the equipment cost $C_3 = 0$, while $C_3 = \$ \, 40,000.00$ in the case of Cloud-RAN. While it is portended that the equipment costs of (both micro and macro) base stations in Cloud-\ac{RAN} would be lower than in \ac{DRAN} (cf. Table \ref{tab:EquipCost}), there is no publicly available information at the moment. Hence, let $\alpha$ denote the extent to which the equipment costs of base stations in the Cloud-\ac{RAN} case are cheaper than in the \ac{DRAN} case, i.e., Cost of a Cloud-\ac{RAN} base station $= \,\alpha \, \times$ Cost of a \ac{DRAN} base station where $0 \le \alpha \le 1$. For example, from Table \ref{tab:EquipCost}, the equipment cost of a \ac{DRAN} macro base station is $ \$ \, 50,000.00$ where as the equipment cost of a Cloud-\ac{RAN} macro base station is $ \$ \, 25,000.00$. Hence, $\alpha = 0.5$. Therefore, in this paper, we consider the equipment cost of base stations in the \ac{DRAN} case to be twice that of the Cloud-RAN case\footnote{Note that this assumption is made based on forecasts by various vendors.} in all subsequent figures with the notable exception of Fig. \ref{Fig:Results.Cost.over.DC.and.RRHcost}(b), where we observe the deployment cost when the equipment cost of a Cloud-\ac{RAN} base station increases to that of a \ac{DRAN} base station.

Fig. \ref{Fig:Results.Cost.over.DC.and.RRHcost_a} is obtained by evaluating equation (\ref{eq:total_cost}) for various data center intensities $\lambda_3$ while keeping all other variables constant. This figure illustrates that deployment costs while employing Cloud-\ac{RAN} data centers are lower than deployment costs in \ac{DRAN} for a wide range of data center intensities. It is also noteworthy that the deployment costs of networks with Cloud-\ac{RAN} are higher than that of \ac{DRAN} only for extremely low and extremely high data center intensities. Considering such values for data center intensities, however, turn out to be quite unrealistic. The extremely high cost for very low data center intensities, i.e., $\lambda_3 < 1$, can be explained as follows. At very low data center intensities, the distances between data centers and backhaul nodes are (on average) very large leading to high infrastructure costs. Furthermore, since all the other parameters such as intensities of base stations, etc. are fixed, the few data centers that are present need to cater to a higher capacity requirement; thereby, leading to higher capacity costs. Hence, a combined effect of high infrastructure as well as capacity costs leads to very high costs for Cloud-\ac{RAN} at $\lambda_3 < 1$. We also observe that at values of $\lambda_3 > 3$, the deployment cost (though still lower than that of \ac{DRAN}) tends to increase. This increase is due to an increase in the total equipment cost (i.e., $\lambda_3 C_3$) which tends to be a major contributing factor and cannot be compensated for by the corresponding reduction in infrastructure and capacity costs. Another salient aspect observed in Fig. \ref{Fig:Results.Cost.over.DC.and.RRHcost_a} is the existence of an ``optimal'' range of data center intensities that minimize the deployment cost of the network while ensuring that user demands are satisfied. Similar behavior was also observed in \cite{ICC'14} and \cite{WiOpt'14} which illustrated the existence of an optimal range of device intensities for minimizing deployment costs while satisfying user requirements.

Fig. \ref{Fig:Results.Cost.over.DC.and.RRHcost_a} also indicates that the decoder quality, represented by the link adaptation offset $\gamma_\text{offset}$, does not significantly affect the deployment cost of a network. While a decoder of poorer quality (viz. a higher $\gamma_\text{offset}$) reduces the data processing complexity required and is, thereby, supposed to reduce the deployment cost, it -- however -- requires more radio access points due to a degradation in performance. These two aspects counteract each other and therefore, results in deployment costs which are not highly dependent on the quality of the decoder utilized. Though the deployment costs appear approximately the same regardless of the quality of the decoder, an aspect which may be critical to the choice of the quality of the decoder to be used is a reduction in processing time. This is because an increase in the link-adaptation offset reduces the number of turbo-decoder iterations and thereby, results in a lower processing latency within the central entity. It is important to note that the effectiveness of Cloud-\ac{RAN} based networks can be attributed to the fact that they are better at adapting to the load (or the traffic) in the network. Contrary to \ac{DRAN} (where each base station is equipped with sufficient resources to handle peak demand), Cloud-\ac{RAN} exploits the computational diversity gain 
(illustrated in Fig.~\ref{fig:complexity_model:complexity_network}) which is obtained by exploiting the temporal- and spatial-traffic fluctuations as well as data processing fluctuations by pooling resources at a central entity (see  \cite{Rost.Talarico.Valenti.TWC.2014} for more details). Therefore, Cloud-\ac{RAN} enables a greater (and improved) utilization of centralized resources.

Fig. \ref{Fig:Results.Cost.over.DC.and.RRHcost_b} illustrates the deployment cost of the network with increasing values of $\alpha$. This figure is utilized to highlight the dependence of deployment cost on the disparity between the equipment costs of Cloud-\ac{RAN} base stations and \ac{DRAN} base stations. We observe that the deployment cost increases as the cost of a Cloud-\ac{RAN} base station approaches that of a \ac{DRAN} base station. Hence, the cost effectiveness of cloud based networks decreases with an increase in the cost of 
base stations, which validates intuition. There are two further observations that can be made from Fig. \ref{Fig:Results.Cost.over.DC.and.RRHcost_b} namely: the deployment cost at $\alpha = 0$ shows the costs incurred for centralized data processing and connecting base stations to data centers; and the difference between the deployment costs of Cloud-\ac{RAN} and \ac{DRAN} at $\alpha = 1$ shows the additional costs required for centralized processing. However, since the cost of Cloud-\ac{RAN} base stations is estimated to be only about half of the cost of \ac{DRAN} base stations (as previously mentioned), we see that Cloud-\ac{RAN} networks are clearly more cost effective than \ac{DRAN} networks.

\begin{figure}
  \centering
    \subfigure[Cost (in millions of dollars) versus user intensity for fiber optic links between backhaul nodes and data centers.]{\includegraphics[width=7cm, height=7cm, keepaspectratio, viewport = 131 348 360 526]{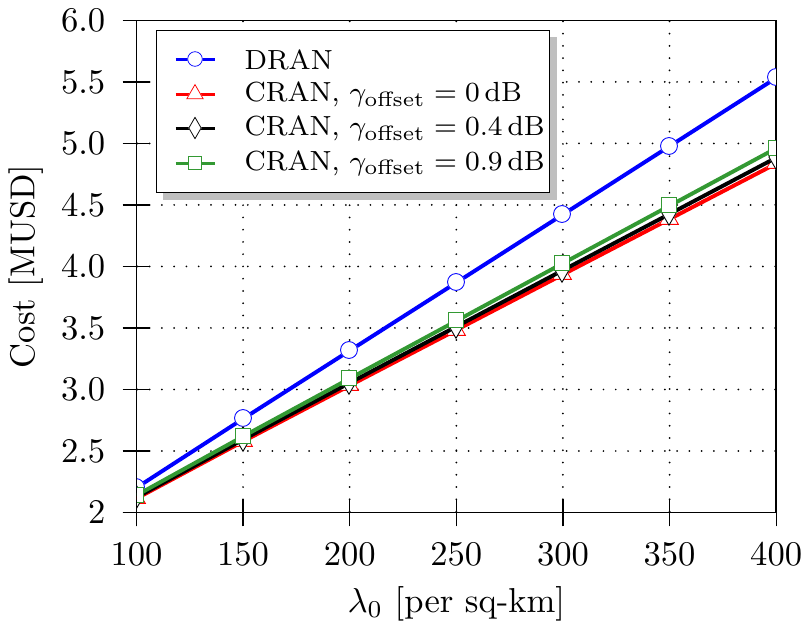} \label{Fig:Results.Cost.over.UserIntensity.and.MWshare_a} }
    \subfigure[Cost (in millions of dollars) versus relative share of microwave backhaul between base station and backhaul layer.]{\includegraphics[width=7cm, height=7cm, keepaspectratio, viewport = 131 348 360 526]{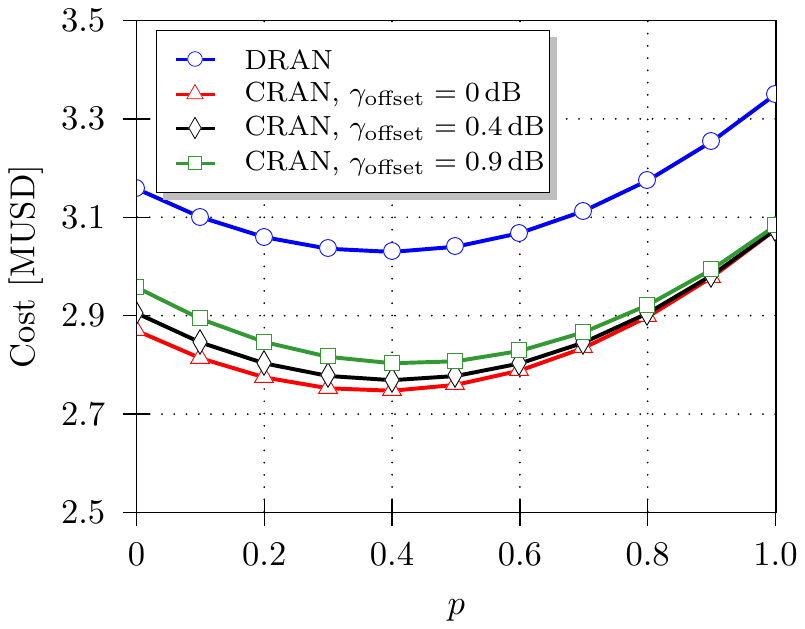} \label{Fig:Results.Cost.over.UserIntensity.and.MWshare_b} }
\caption{Illustrating the variation in deployment cost with increasing user intensity and an increase in the share of the microwave backhaul.}
\vspace{-5mm}
\label{Fig:Results.Cost.over.UserIntensity.and.MWshare}
\end{figure}

Next, consider Fig. \ref{Fig:Results.Cost.over.UserIntensity.and.MWshare} which shows two more examples of how the parameterization chosen impacts the deployment costs of a Cloud-\ac{RAN} network, i.e., the traffic demand represented by user intensity $\lambda_0$ and the share of microwave backhaul represented by $p$ where $p=0$ represents
deployments with only fiber optic backhaul and $p=1$ represents the case with only microwave backhaul technologies. Fig. \ref{Fig:Results.Cost.over.UserIntensity.and.MWshare_a} highlights the dependence of deployment cost on the user intensity $\lambda_0$ and we see that 
the cost effectiveness of Cloud-\ac{RAN} networks increases with an increase in $\lambda_0$. This is because Cloud-\ac{RAN} networks are better at adapting to the network load than \ac{DRAN} networks though the base station intensity (and the corresponding backhaul intensity) as well as connectivity requirements increase in both Cloud-\ac{RAN} and \ac{DRAN} networks.

Furthermore, Fig. \ref{Fig:Results.Cost.over.UserIntensity.and.MWshare_b} shows that the deployment type whose deployment cost is the lowest is one where both microwave and fiber optic backhaul technologies exist (viz. $0 < p < 1$) rather than deployments with just one technology or the other (viz. $p = 0 \vee p = 1$ ). We also observe that if $p = 1$, i.e., only microwave backhaul deployment, there is no perceivable difference in cost between different decoder configurations, but the overall deployment costs, however, are lower when $p = 0$. It is also interesting to note that for $p = 0$, i.e., only fiber optic backhaul deployment, the deployment cost of a network using a decoder of lower quality is higher than that of a network using a decoder of higher quality. This is due to the fact that using a lower quality decoder necessitates an increase in base station intensity in order to compensate for the performance degradation. This increase in base station intensity is then accompanied by a proportionate increase in the overall cost of deployment.

\begin{figure}[!t]
  \centering
    \subfigure[Cost (in millions of dollars) versus $\sigma^2$ for fiber optic links between backhaul nodes and data centers, i.e., $p = 0$.]{\includegraphics[trim=0cm 7cm 0cm 7cm,clip=true,width=0.78\linewidth]{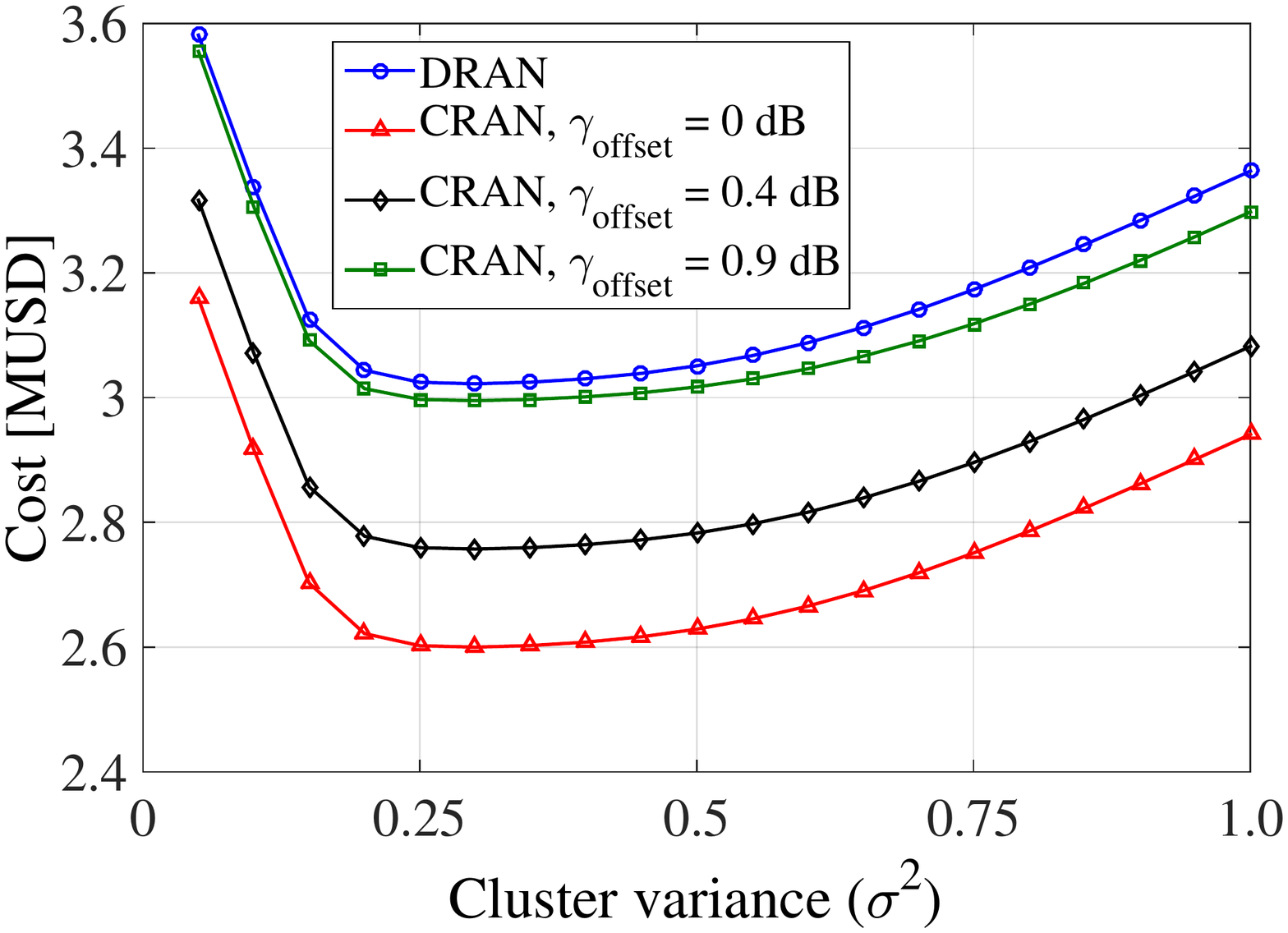} \label{Fig:Cost_CRAN_DRAN_vs_sigma_OF} }
    \subfigure[Cost (in millions of dollars) versus $\sigma^2$ for both microwave and fiber optic links between backhaul nodes and data centers, i.e., $p = 0.5$.]{\includegraphics[trim=0cm 7cm 0cm 7cm,clip=true,width=0.78\linewidth]{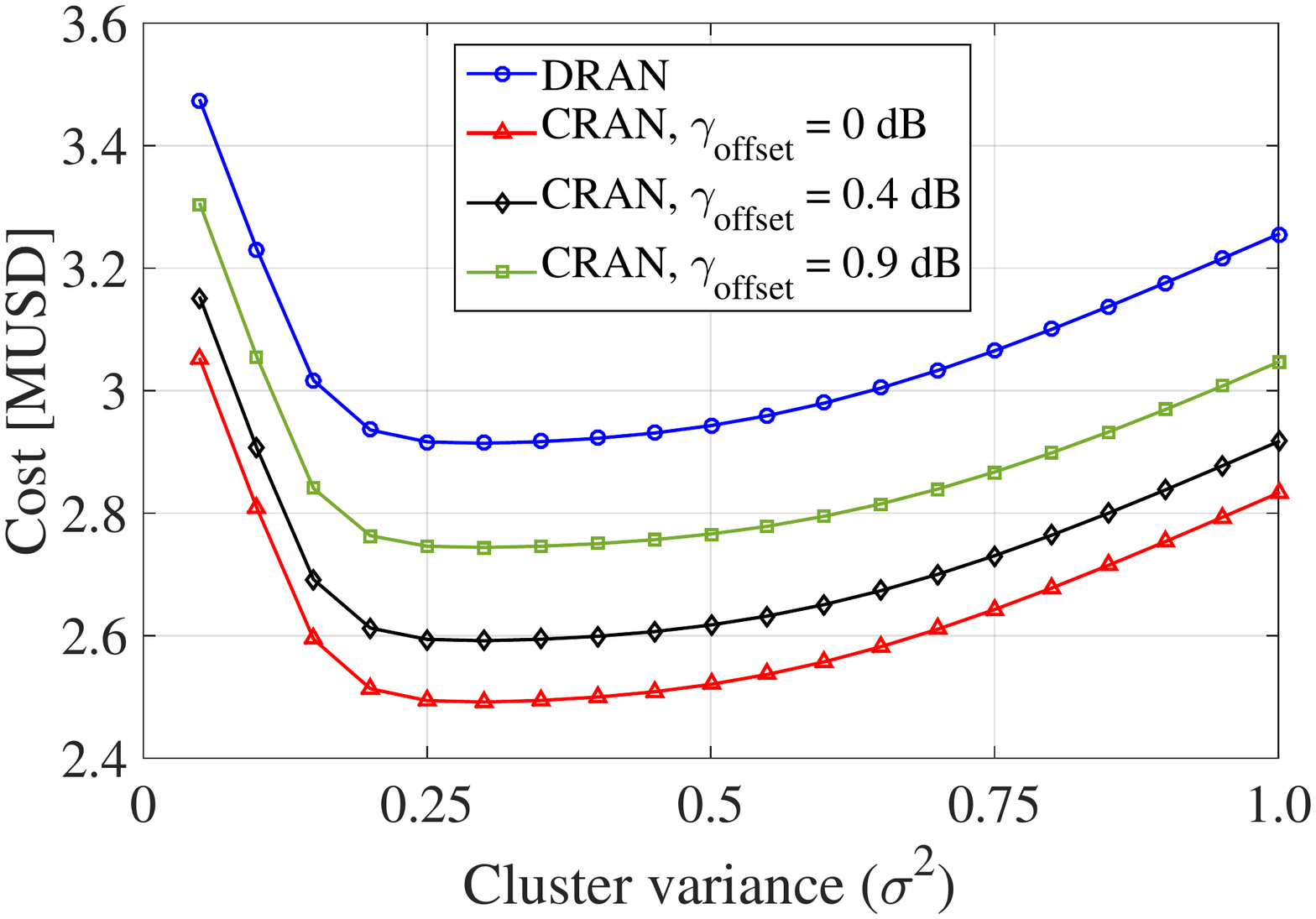} \label{Fig:Cost_CRAN_DRAN_vs_sigma} }
    \subfigure[Cost (in millions of dollars) versus $\sigma^2$ for microwave links between backhaul nodes and data centers, i.e., $p = 1$.]{\includegraphics[trim=0cm 7cm 0cm 7cm,clip=true,width=0.78\linewidth]{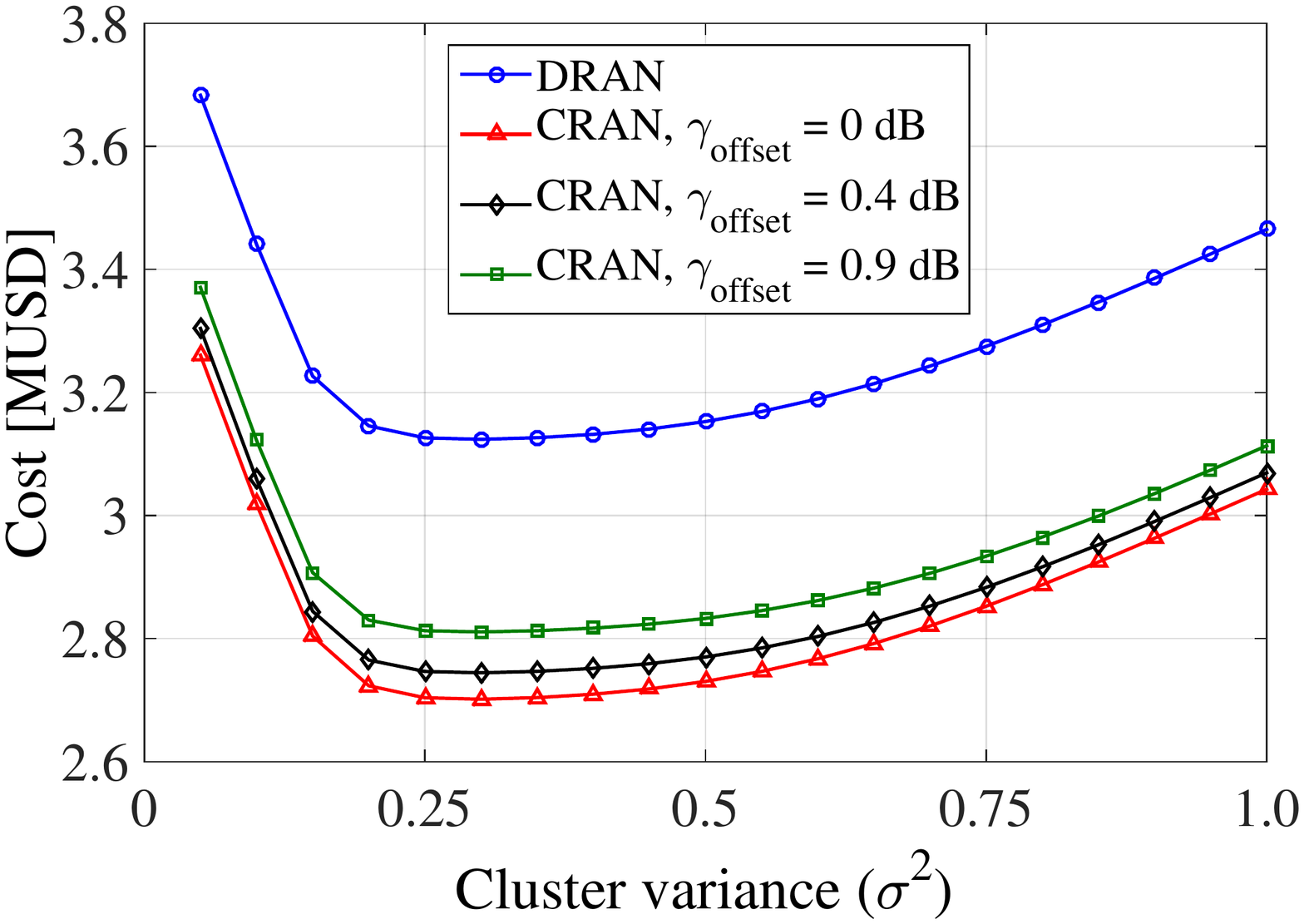} \label{Fig:Cost_CRAN_DRAN_vs_sigma_MW} }
\caption{Illustrating the variation in cost with increasing cluster variance $(\sigma^2)$. Note that all other parameters are fixed and data center intensity $\lambda_3 = 3/\text{km}^2$.}
\vspace{-5mm}
\label{Fig:Results.Cost_CRAN_DRAN_vs_sigma}
\end{figure}

\begin{figure*}[!thb]
\begin{align}
\label{App:eq:Initial_cost_formulation}
\mathcal{C}_{\Phi_3} &= \mathbb{E}_{\Phi_3}^o \Bigg[\sum\limits_{z \in \Phi_2 \cap V_o(\Phi_3)}\!\!\bigg(\!C_2 \! + \! \mathcal{N}_{z} \left( A'_{2,3} \|z\|^{\beta_{2,3}}
 \!+\! A''_{2,3} \right)\! + \! B_{2,3} \|z\|^{\theta_{2,3}} \! + \!\!\!
 \sum\limits_{y \in \Phi_1 \cap V_z(\Phi_2)}\!\Big\{ C_1\! +\! \mathcal{N}_{y} A_{1,2} \|y - z\|^{\beta_{1,2}}\! +\! B_{1,2} \|y - z\|^{\theta_{1,2}} \, \nonumber \\
& \hspace{3.5cm} {+}\sum\limits_{x \in \Phi_0 \cap V_{y}(\Phi_1)} \! \left( A_{0,1}\|x - y - z\|^{\beta_{0,1}} + \, B_{0,1}\|x - y - z\|^{\theta_{0,1}} \right) \Big\} \bigg) \Bigg] \\
\label{eq:IndivCost}
\mathcal{C}_{\Phi_3} &= \frac{\lambda_2}{\lambda_3} \Bigg[ C_2 + \mathbb{E}_{\Phi_2}^o \left[\mathcal{N}_o \left( A'_{2,3} \|z_o\|^{\beta_{2,3}} + A''_{2,3} \right) \right] + \mathbb{E}_{\Phi_2}^o \left[B_{2,3} \|z_o\|^{\theta_{2,3}} \right] \, + \nonumber \\
& \hspace{0.5cm} \mathbb{E}_{\Phi_2}^o \bigg[ \sum\limits_{y \in \Phi_1 \cap V_o(\Phi_2)} \Big\{ C_1 + \mathcal{N}_{y} A_{1,2} \|y\|^{\beta_{1,2}} + B_{1,2} \|y\|^{\theta_{1,2}} \, + 
 \sum\limits_{x \in \Phi_0 \cap V_{y}(\Phi_1)} \! \left( A_{0,1}\|x - y\|^{\beta_{0,1}} + \, B_{0,1}\|x - y\|^{\theta_{0,1}} \right) \Big\} \bigg] \Bigg]
\end{align}
\hrulefill
\end{figure*}

Finally, Fig. \ref{Fig:Results.Cost_CRAN_DRAN_vs_sigma} examines the effect of the cluster variance (i.e., the spread of the micro base stations around the macro base stations) on the deployment cost of the network. The data center intensity $\lambda_3$ is chosen to be $3/\text{km}^2$ based on Fig. \ref{Fig:Results.Cost.over.DC.and.RRHcost_a}(viz. $\sigma^2 = 0.5$). Fig. \ref{Fig:Results.Cost_CRAN_DRAN_vs_sigma} examines three different scenarios in its sub-figures. The first, Fig. \ref{Fig:Cost_CRAN_DRAN_vs_sigma_OF}, shows variations in deployment costs with increasing cluster variance when only fiber optic backhauling is used. Then, Fig. \ref{Fig:Cost_CRAN_DRAN_vs_sigma} considers variations in the deployment costs of a network (which has a mix of both fiber optic and microwave backhaul technologies) with increasing cluster variance. Since, Fig. \ref{Fig:Results.Cost.over.UserIntensity.and.MWshare_b} shows that the deployment cost is minimum at approximately $p = 0.5 \,$, we chose the same value in Fig. \ref{Fig:Cost_CRAN_DRAN_vs_sigma}. Lastly, Fig.~\ref{Fig:Cost_CRAN_DRAN_vs_sigma_MW} illustrates the variations in deployment costs with increasing cluster variance when only microwave backhauling is used. By comparing the deployment costs in all the three sub-figures above regardless of whether we consider Cloud-\ac{RAN} or \ac{DRAN}, we see that the overall deployment costs are minimized when both types of backhaul technologies coexist. Hence, corroborating Fig. \ref{Fig:Results.Cost.over.UserIntensity.and.MWshare_b}. Another interesting observation is the fact that, unlike Fig. \ref{Fig:Results.Cost.over.DC.and.RRHcost}  and Fig. \ref{Fig:Results.Cost.over.UserIntensity.and.MWshare}, the deployment cost curves for the various offsets in the Cloud-\ac{RAN} case are much farther apart. This behavior indicates that the extent of spread of the micro base stations around the macro base stations has a significant impact on the deployment costs of Cloud-\ac{RAN} networks.

\section{Conclusions}\label{End}
This paper analyzes the cost effectiveness of Cloud-\ac{RAN} based heterogeneous networks against a distributed implementation of LTE heterogeneous networks. The main result of the work is a theoretic framework which helps compute the deployment cost of a network. A complexity model, which provides information about processing costs and its dependence on base station intensities, is used as an input along with several other base costs to compute the deployment cost of a distributed heterogeneous LTE network as well as Cloud-\ac{RAN} based heterogeneous networks. We show that Cloud-\ac{RAN} based heterogeneous networks cost less than standard LTE deployments and are, therefore, more cost effective than distributed networks. The findings also reveal that deploying a mix of backhaul technologies is more cost effective than using just one type of technology. Finally, it is also observed that the spread of the micro base stations around the macro base stations has a significant impact on the deployment cost of Cloud-\ac{RAN} networks. 

\appendix

\section{Proof of the Theorem}\label{Theorem_Proof}

\begin{proof}
The average cost of deploying a data center, $\mathcal{C}_{\Phi_3}$, can be defined as equation (\ref{App:eq:Initial_cost_formulation}) where $\mathbb{E}_{\Phi_3}^o\left[\cdot \right]$ is the Palm expectation with respect to the process $\Phi_3$. Separating the terms after using the exchange formula of Neveu (see \cite{SIG'08}), substituting the individual values of $A_{2,3}$, and taking into account that $A''_{2,3}$ is independent of distance results in equation (\ref{eq:IndivCost}) where the point of observation is shifted to the origin `$o$' and $z_o$ is the point of the point process $\Phi_2$ closest to `$o$'.

Equation (\ref{eq:IndivCost}) can be solved using a sort of an iterative process. First, consider the second and third terms of the RHS. Since the point processes in each layer are independent of the others, we can write the second and third terms of the RHS of equation (\ref{eq:IndivCost}) as $\mathbb{E}_{\Phi_2}^o \left[\mathcal{N}_o\right]  \, \mathbb{E}_{\Phi_2}^o \left[A'_{2,3} \|z_o\|^{\beta_{2,3}} \right] + \mathbb{E}_{\Phi_2}^o \left[\mathcal{N}_o\right]  \, \mathbb{E}_{\Phi_2}^o \left[A''_{2,3} \right] $. Furthermore, from \cite{FB'09}, we get $\mathbb{E}_{\Phi_2}^o \left[\mathcal{N}_o\right] = \frac{\lambda_0}{\lambda_2}$ and since $A''_{2,3}$ is a constant, $\mathbb{E}_{\Phi_2}^o \left[\mathcal{N}_o\right]  \, \mathbb{E}_{\Phi_2}^o \left[A''_{2,3} \right] = \frac{\lambda_0}{\lambda_2} A''_{2,3}$. The terms $\mathbb{E}_{\Phi_2}^o \left[A'_{2,3} \|z_o\|^{\beta_{2,3}} \right]$ and $\mathbb{E}_{\Phi_2}^o \left[B_{2,3} \|z_o\|^{\theta_{2,3}} \right]$ can (both) be simplified as shown below.
\begin{align}
\label{eq:CostTerm2}
& \mathbb{E}_{\Phi_2}^o \left[A'_{2,3} \|z_o\|^{\beta_{2,3}} \right] \nonumber \\[5pt]
& \stackrel{(a)}{=} A'_{2,3} \mathbb{E} \int\limits_{\mathbb{R}^2} \|a\|^{\beta_{2,3}} \mathbf{1}\left( \Phi_2 \left( b(o,\|a\|) = 0 \right) \right) \Phi_3(\mathrm{d}a) \nonumber \\
& \stackrel{(b)}{=} A'_{2,3} \, \lambda_3 \int\limits_{\mathbb{R}^2} \|a\|^{\beta_{2,3}} \mathbb{P}_3^o\left( b(-a,\|a\|) \right) \mathrm{d}a, 
\end{align}
where $b(o,\|a\|)$ is an open ball of radius $\|a\|$ centered at $o$. Here, equality $(a)$ is due to the independence of the point processes, $\mathbf{1}(\cdot)$ is the indicator function, and $\Phi_3(\cdot)$ is now a stationary counting measure on $\mathbb{R}^2$. The equality $(b)$ is due to the Refined Campbell theorem (see \cite{SIG'08}) where $\mathbb{P}_3^o (\cdot)$ is the Palm distribution with respect to the point process $\Phi_3$. Since $\Phi_3$ is a homogeneous Poisson point process, we get
\begin{align}
\label{eq:A23}
\mathbb{E}_{\Phi_2}^o \left[A'_{2,3} \|z_o\|^{\beta_{2,3}} \right] 
& \stackrel{(c)}{=} 2\pi A'_{2,3} \lambda_3 \int\limits_{\mathbb{R}^+} r^{\beta_{2,3} + 1} \exp \left(-\pi \lambda_3 r^2 \right)\! \mathrm{d}r \nonumber \\
& =  A'_{2,3} \frac{\Gamma\left(\frac{\beta_{2,3}}{2} + 1 \right)}{\left(\pi \lambda_3 \right)^{\beta_{2,3}/2}},
\end{align}
where the RHS of equality $(c)$ is in radial coordinates, which (upon computation) results in the expression with the Gamma function. The fourth term of equation (\ref{eq:IndivCost}) can also be found along the same lines and results in 
\begin{equation}
\label{eq:B23}
\mathbb{E}_{\Phi_2}^o \left[B_{2,3} \|z_o\|^{\theta_{2,3}} \right]  =  B_{2,3} \frac{\Gamma\left(\frac{\theta_{2,3}}{2} + 1 \right)}{\left(\pi \lambda_3 \right)^{\theta_{2,3}/2}}.
\end{equation}
Consider the inner Palm expectation term, (i.e., the fifth term) of equation (\ref{eq:IndivCost}). Using the exchange formula of Neveu \cite{SIG'08}, we obtain equation (\ref{eq:IndivCost_PCP})
\begin{figure*}[t]
\begin{align}
\label{eq:IndivCost_PCP}
& \mathbb{E}_{\Phi_2}^o \Bigg[ \! \sum\limits_{y \in \Phi_1 \cap V_o(\Phi_2)} \!\! \Big\{ C_1 + \mathcal{N}_{y} A_{1,2} \|y\|^{\beta_{1,2}} + B_{1,2} \|y\|^{\theta_{1,2}} + \!\!\!\! \sum\limits_{x \in \Phi_0 \cap V_{y}(\Phi_1)} \!\!\!\!\! \left( A_{0,1}\|x - y\|^{\beta_{0,1}} + B_{0,1}\|x - y\|^{\theta_{0,1}} \! \right) \! \Big\} \! \Bigg] \nonumber \\[5pt]
& \hspace{0.5cm} = \frac{\lambda_1}{\lambda_2} \bigg[ C_1 + \mathbb{E}_{\Phi_1}^o \left[\mathcal{N}_o A_{1,2} \|y_o\|^{\beta_{1,2}} \right] + \mathbb{E}_{\Phi_1}^o \left[B_{1,2} \|y_o\|^{\theta_{1,2}} \right] \, + 
\mathbb{E}_{\Phi_1}^o \Big[\rule{0cm}{0.65 cm} \sum\limits_{x \in \Phi_0 \cap V_{o}(\Phi_1)} \!\!\! \left( A_{0,1}\|x\|^{\beta_{0,1}} + B_{0,1}\|x\|^{\theta_{0,1}} \right) \Big] \bigg] 
\end{align}
\hrulefill
\end{figure*}
where (as before) the point of observation is shifted to the origin `$o$' and $y_o$ is the point of the point process $\Phi_1$ closest to `$o$'. 
Since the point processes in each layer are independent and $\mathbb{E}_{\Phi_1}^o \left[\mathcal{N}_o\right] = \frac{\lambda_0}{\lambda_1}$, \cite{FB'09}. Hence, we can write the second term of the RHS of equation (\ref{eq:IndivCost_PCP}) as $\frac{\lambda_0}{\lambda_1} \, \mathbb{E}_{\Phi_1}^o \left[A_{1,2} \|y_o\|^{\beta_{1,2}} \right]$. The terms $\mathbb{E}_{\Phi_1}^o \left[A_{1,2} \|y_o\|^{\beta_{1,2}} \right]$ and $\mathbb{E}_{\Phi_1}^o \left[B_{1,2} \|y_o\|^{\theta_{1,2}} \right]$ can be written as
\begin{align}
\label{eq:CostTerm2_PCP}
& \mathbb{E}_{\Phi_1}^o \left[A_{1,2} \|y_o\|^{\beta_{1,2}} \right] \nonumber \\
& = A_{1,2} \mathbb{E} \int\limits_{\mathbb{R}^2} \|a\|^{\beta_{1,2}} \mathbf{1}\left( \Phi_2 \left( b(o,\|a\|) = 0 \right) \right) \Phi_2(\mathrm{d}a) \nonumber \\
& = A_{1,2} \, \lambda_2 \int\limits_{\mathbb{R}^2} \|a\|^{\beta_{1,2}} \mathbb{P}_2^o\left( b(-a,\|a\|) \right) \mathrm{d}a 
\end{align}
using the independence of the point processes and the Refined Campbell Theorem. Recall that $\Phi_2$ is a stationary mixed Poisson process. Hence, its Palm distribution is given by
\begin{align}
\label{eq:Palm_2}
& \mathbb{P}_2^o\left( b(-a,\|a\|) \right) \nonumber \\[5pt]
& \hspace{1cm} = \frac{1}{\lambda_2} \bigg[ p \, \lambda_{2\text{MW}} \mathbb{P}_{2\text{MW}}^o\left( b(-a,\|a\|) \right) \, + \nonumber \\
& \hspace{3cm} (1-p) \lambda_{2\text{OF}} \mathbb{P}_{2\text{OF}}^o\left( b(-a,\|a\|) \right) \bigg] \nonumber \\
& \hspace{1cm} = \frac{1}{\lambda_2}\bigg[ p \, \lambda_{2\text{MW}} \exp\big( -\pi \lambda_{2\text{MW}} \|a\|^2 \big) \, + \nonumber \\
& \hspace{2.6cm} (1-p) \lambda_{2\text{OF}} \exp\big( -\pi \lambda_{2\text{OF}} \|a\|^2 \big) \bigg].
\end{align}
Hence, substituting equation (\ref{eq:Palm_2}) in (\ref{eq:CostTerm2_PCP}) and integrating, we obtain 
\begin{align}
\label{eq:Psi_1}
& \Psi_1\big(A_{1,2}, \beta_{1,2}, \lambda_{2\text{MW}}, \lambda_{2\text{OF}}, p \big) \equiv \mathbb{E}_{\Phi_1}^o \left[A_{1,2} \|y_o\|^{\beta_{1,2}} \right] \nonumber \\[5pt]
& \hspace{5mm} = A_{1,2} \Bigg[ \frac{p \, \Gamma(\frac{\beta_{1,2}}{2}+1)}{(\pi\lambda_{2\text{MW}})^{\beta_{1,2}/2}} + \frac{(1-p) \, \Gamma(\frac{\beta_{1,2}}{2}+1)}{(\pi\lambda_{2\text{OF}})^{\beta_{1,2}/2}} \Bigg]\\
\label{eq:Psi_2}
& \Psi_2\big(B_{1,2}, \theta_{1,2}, \lambda_{2\text{MW}}, \lambda_{2\text{OF}}, p \big) \equiv \mathbb{E}_{\Phi_1}^o \left[B_{1,2} \|y_o\|^{\theta_{1,2}} \right] \nonumber \\[5pt]
& \hspace{5mm} = B_{1,2} \Bigg[ \frac{p \, \Gamma(\frac{\theta_{1,2}}{2}+1)}{(\pi\lambda_{2\text{MW}})^{\theta_{1,2}/2}} + \frac{(1-p) \, \Gamma(\frac{\theta_{1,2}}{2}+1)}{(\pi\lambda_{2\text{OF}})^{\theta_{1,2}/2}} \Bigg].
\end{align}
Finally, the Palm expectation of the last term of the RHS of equation (\ref{eq:IndivCost_PCP}) can be simplified using the exchange formula of Neveu:
\begin{align}
\label{eq:CostTerms34_PCP}
& \mathbb{E}_{\Phi_1}^o \bigg[ \sum\limits_{x \in \Phi_0 \cap V_{o}(\Phi_1)} \left( A_{0,1}\|x\|^{\beta_{0,1}} + B_{0,1}\|x\|^{\theta_{0,1}} \right) \bigg] \nonumber \\
& = \frac{\lambda_0}{\lambda_1} \Big[  \mathbb{E}_{\Phi_0}^o \left[ A_{0,1} \|a_o\|^{\beta_{0,1}} \right] + \mathbb{E}_{\Phi_0}^o \left[B_{0,1} \|a_o\|^{\theta_{0,1}} \right] \Big],
\end{align}
where $\|a_o\|$ is the effective distance between the base station at $o$ and a user. With the use of the Refined Campbell theorem, the first term on the RHS of equation (\ref{eq:CostTerms34_PCP}) is
\begin{align}
\label{eq:CostTerm3_PCP}
\mathbb{E}_{\Phi_0}^o \left[A_{0,1} \|a_o\|^{\beta_{0,1}} \right] & \nonumber \\
& \hspace{-1.5cm} = A_{0,1} \, \lambda_1 \int\limits_{\mathbb{R}^2} \|a\|^{\beta_{0,1}} \mathbb{P}_1^o\left( b(-a,\|a\|) \right) \mathrm{d}a. 
\end{align}
Since $\Phi_1$ is a Poisson cluster process, finding its Palm distribution $\mathbb{P}_1^o(\cdot)$ is slightly more complicated. Using the $J$-function (see \cite{Lieshout'96}) and the ``empty space function", $F$ (see \cite{Moeller'02}), it is written as
\begin{equation}
\label{eq:Palm_Phi1}
\mathbb{P}_1^o(b(o,R)) = 1 - \left[1 - F_{\Phi_1}(R)\right]J_{\Phi_1}(R),
\end{equation}
where $R$ is the random distance from $o$ to the nearest point in $\Phi_1$ (due to the stationarity of $\Phi_1$). Note that for a realization $r$ of the distance $R$, the Palm distribution can be obtained by  taking the derivative of equation (\ref{eq:Palm_Phi1}) with respect to $r$. The $J$-function of $J_{\Phi_1}(R)$, is given by 
\begin{align*}
& J_{\Phi_1}(R) = J_{\Phi_{1c} \cup \Phi_{1m}}(R) \\[5pt]
& = \frac{\lambda_{1c}}{\lambda_{1c}+\lambda_{1c}\lambda_{1m}}J_{\Phi_{1c}}(R) + \frac{\lambda_{1c}\lambda_{1m}}{\lambda_{1c}+\lambda_{1c}\lambda_{1m}}J_{\Phi_{1m}}(R),
\end{align*}
since the processes $\Phi_{1c}$ and $\Phi_{1m}$ are independent stationary point processes (see \cite{Lieshout'96} and \cite{Baddeley'07} for details). As shown in \cite{Lieshout'96}, since $\Phi_{1c}$ is a stationary Poisson process, $J_{\Phi_{1c}}(R) = 1$ and $J_{\Phi_{1m}}(R)$ can be derived as 
\begin{equation*}
J_{\Phi_{1m}}(R) = \bigintssss\limits_{\mathbb{R}^2} \hspace{-2mm} f(x) \exp\Big(- \hspace{-2mm} \int\limits_{\|y\| \le R} \hspace{-2mm} \lambda_{1m} f(y+x) \mathrm{d}y \Big) \mathrm{d}x,
\end{equation*}
from the general expression for stationary Cox processes provided in \cite{Moeller'02}. Hence, the $J$-function can be written as
\begin{align}
\label{eq:J-function}
& J_{\Phi_1}(R) = \frac{\lambda_{1c}}{\lambda_{1c}+\lambda_{1c}\lambda_{1m}} \, + \nonumber \\[8pt] 
& \frac{\lambda_{1c}\lambda_{1m}}{\lambda_{1c}+\lambda_{1c}\lambda_{1m}}\bigg[\bigintssss\limits_{\mathbb{R}^2} \hspace{-2mm} f(x) \exp\Big( \hspace{-2mm} - \hspace{-4mm} \int\limits_{\|y\| \le R} \hspace{-3mm} \lambda_{1m} f(y+x) \mathrm{d}y \Big) \mathrm{d}x \bigg].
\end{align}
Then, recalling that $\left[1-F_{\Phi_1}(R)\right]$ is the \textit{void probability} (see \cite{Stoyan'95}), we get
\begin{align}
\label{eq:Void_Prob_Phi1}
1 - F_{\Phi_1}(R) =& \exp \Bigg( \hspace{-2mm} -\lambda_{1c} \hspace{-2mm} \bigintsss\limits_{\mathbb{R}^2} \hspace{-1mm} \bigg[1 \, - \, \mathbf{1}\left(x \notin  b(o,R)\right) \times \nonumber \\
& \exp\Big( \hspace{-2mm} -\lambda_{1m} \hspace{-2mm} \int\limits_{ b(o,R)} \hspace{-2mm} f(y-x) \, \mathrm{d}y \Big) \bigg] \mathrm{d}x \Bigg).
\end{align}
Hence, the Palm distribution can be found by substituting equations (\ref{eq:J-function}) and (\ref{eq:Void_Prob_Phi1}) in equation~(\ref{eq:Palm_Phi1}). Therefore, for a realization $R=r$, equation (\ref{eq:CostTerm3_PCP}) becomes 
\begin{align}
\label{eq:Psi_3}
& \Psi_3\big(A_{0,1}, \beta_{0,1}, \lambda_{1c}, \lambda_{1m}, f(\cdot), \sigma^2 \big) \equiv \mathbb{E}_{\Phi_0}^o \left[A_{0,1} \|a_o\|^{\beta_{0,1}} \right] \nonumber \\[5pt]
& = A_{0,1} \, \lambda_1 \int\limits_{\mathbb{R}} r^{\beta_{0,1}} \mathbb{P}_1^o\left( b(o,r) \right) \mathrm{d}r \\
\label{eq:Psi_4}
& \Psi_4\big(B_{0,1}, \theta_{0,1}, \lambda_{1c}, \lambda_{1m}, f(\cdot), \sigma^2 \big) \equiv \mathbb{E}_{\Phi_0}^o \left[B_{0,1} \|a_o\|^{\theta_{0,1}} \right] \nonumber \\[5pt]
& = B_{0,1} \, \lambda_1 \int\limits_{\mathbb{R}} r^{\theta_{0,1}} \mathbb{P}_1^o\left( b(o,r) \right) \mathrm{d}r.
\end{align}
Then, substitute equations (\ref{eq:Psi_3}) and (\ref{eq:Psi_4}) in equation (\ref{eq:CostTerms34_PCP}). Finally, substituting equations (\ref{eq:Psi_1}), (\ref{eq:Psi_2}), and (\ref{eq:CostTerms34_PCP}) in equation (\ref{eq:IndivCost_PCP}) and in turn substituting equations (\ref{eq:A23}), (\ref{eq:B23}), and (\ref{eq:IndivCost_PCP}) in equation (\ref{eq:IndivCost}) results in equation (\ref{eq:theorem1}). Note that equations (\ref{eq:Psi_3}) and (\ref{eq:Psi_4}) can be evaluated easily using numerical integration. 
\end{proof}


\bibliographystyle{IEEEtran}
\bibliography{JSAC}

\vspace*{-3\baselineskip}
\begin{IEEEbiography}[{\includegraphics[width=1in,height=1.25in,clip,keepaspectratio]{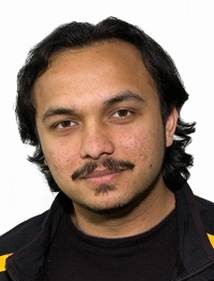}}]{Dr. Vinay Suryaprakash}
received his doctorate from the Technische Universit\"{a}t Dresden, Germany under the supervision of Prof. Gerhard Fettweis in 2014 and his Master of Science from the University of Southern California, Los Angeles in 2007. His research focuses on using stochastic geometry for the system level analysis of wireless networks. In 2013, he was nominated as one of the six finalists of the Qualcomm Innovation Fellowship 2013 from contestants all across Europe.
\end{IEEEbiography}
\vspace*{-2\baselineskip}
\begin{IEEEbiography}[{\includegraphics[width=1in,height=1.25in,clip,keepaspectratio]{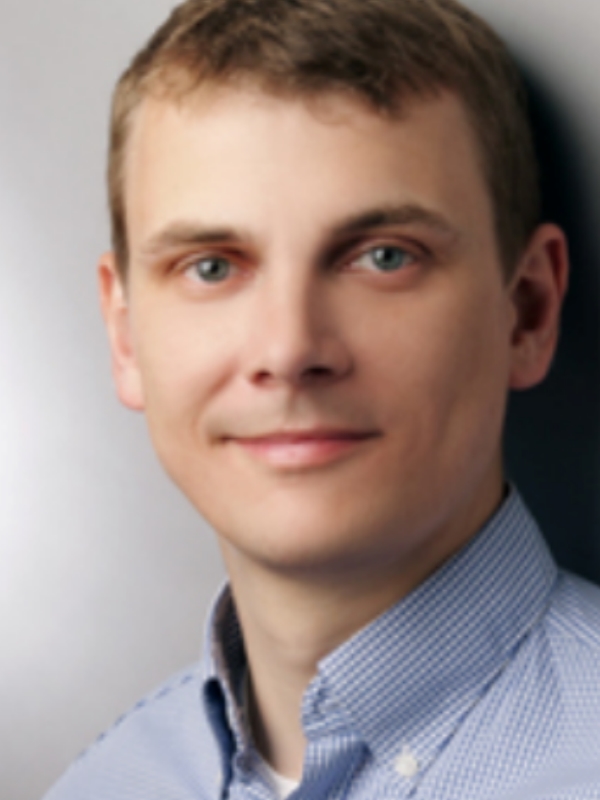}}]{Dr. Peter Rost} received his Ph.D. degree from Technische Universität Dresden, Germany, in 2009 (supervised by Prof. Gerhard Fettweis) and his M.Sc. from University of Stuttgart, Germany, in 2005. Since April 2010, he has been a member of the Wireless and Backhaul Networks group at NEC Laboratories Europe, where he is a Senior Researcher involved in business unit projects, 3GPP RAN2, and the EU FP7 project iJOIN, which he currently leads as Technical Manager (www.ict-ijoin.eu). Peter is member of IEEE ComSoc GITC, IEEE Online GreenComm Steering Committee, and the Executive Editorial Committee of IEEE Transactions of Wireless Communications as well as a member of VDE and ITG expert committee “Information and System Theory”.
\end{IEEEbiography}
\vspace*{-2\baselineskip}
\begin{IEEEbiography}[{\includegraphics[width=1in,height=1.25in,clip,keepaspectratio]{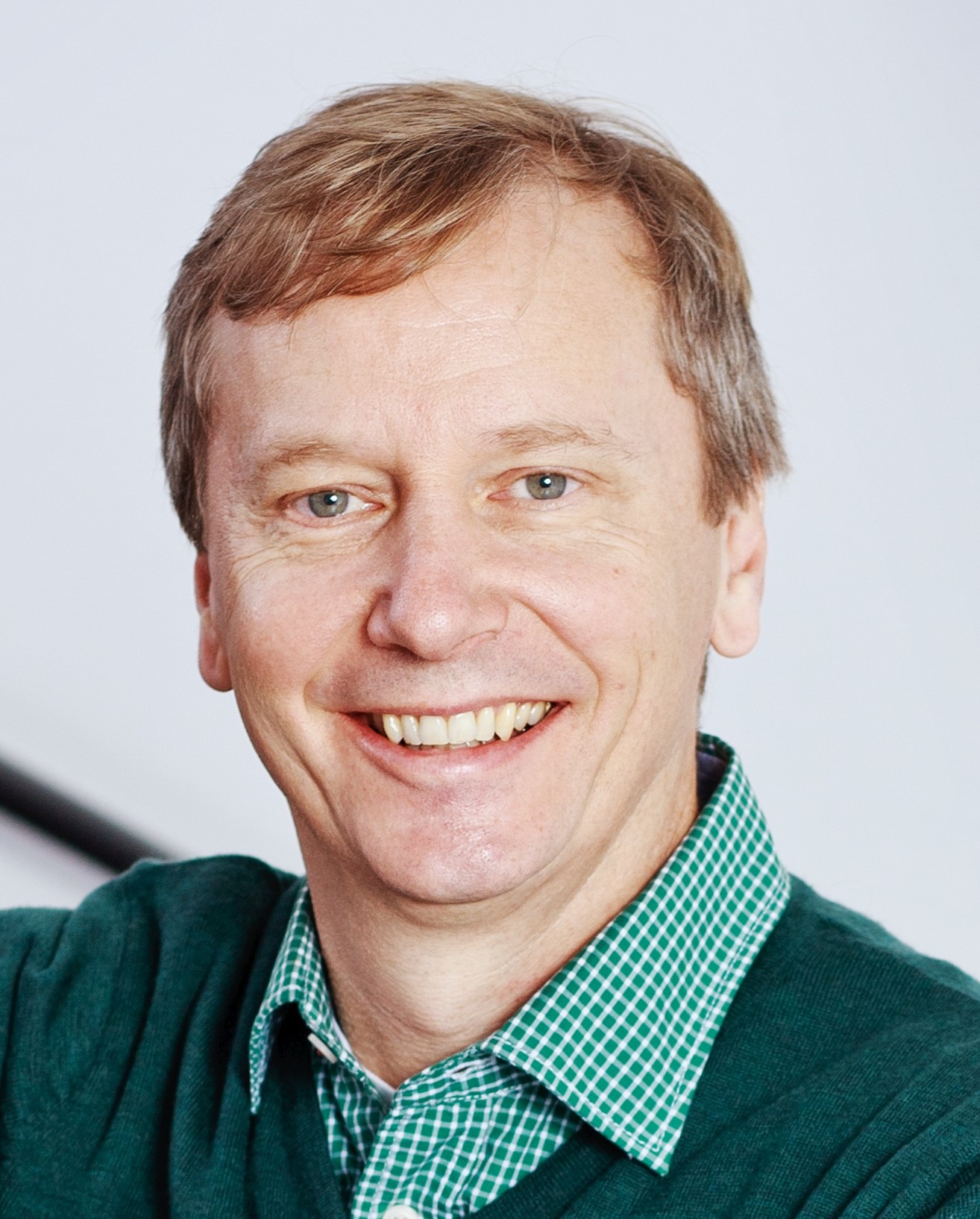}}]{Prof. Dr.-Ing. Dr. h.c. Gerhard Fettweis} earned his Dipl.-Ing. and Ph.D. degrees from Aachen University of Technology (RWTH) in Germany. From 1990 to 1991, he was a visiting scientist at the IBM Almaden Research Center in San Jose, CA, working on signal processing for disk drives. From 1991 to 1994, he was with TCSI Inc., Berkeley, CA, responsible for signal processor developments. Since September 1994 he holds the Vodafone Chair at Technische Universit\"{a}t Dresden, Germany. In 2012, he received an Honorary Doctorate from Tampere University. 
He is also a well known serial entrepreneur who has co-founded 11 start-ups to date. He has been an elected member of the IEEE Solid State Circuits Society's Board (Administrative Committee) since 1999, and he is also an elected member of the IEEE Fellow Committee. 
Since 2011, he has been the Coordinator of the DFG Collaborative Research Center SFB 912 ``Highly Adaptive Energy Efficient Computing'' and from 2012, he has been the Director and Scientific Coordinator of the WR/DFG German Cluster of Excellence ``Center for Advancing Electronics Dresden''. Apart from which, he serves on company supervisory boards and on industrial as well as research institutes’ advisory committees.
\end{IEEEbiography}

\end{document}